\documentclass[
showpacs,
preprintnumbers,
reprint,
amsmath,amssymb,
10pt
]
{revtex4-1}

\usepackage{graphicx}
\usepackage{dcolumn}
\usepackage{bm}

\usepackage[english]{babel} 

\renewcommand{\vec}{\bm} 
\newcommand{\mat}{\bm}   
\newcommand{\uvec}[1]{{\bm{\hat{#1}}}} 

\begin{document}

\preprint{APS/123-QED}

\title{Colloidal suspensions of C-particles:\\ Entanglement, percolation and microrheology\/}

\author{Christian Hoell}
\email[Electronic address: ]{choell@thphy.uni-duesseldorf.de}
\affiliation{%
Institut f\"ur Theoretische Physik II: Weiche Materie, 
Heinrich-Heine-Universit\"at D\"usseldorf\\
Universit\"atsstrasse 1, D-40225 D\"usseldorf, Germany 
}
\author{Hartmut L\"owen}
\email[Electronic address: ]{hlowen@thphy.uni-duesseldorf.de}
\affiliation{%
Institut f\"ur Theoretische Physik II: Weiche Materie, 
Heinrich-Heine-Universit\"at D\"usseldorf\\
Universit\"atsstrasse 1, D-40225 D\"usseldorf, Germany 
}

\date{\today}

\begin{abstract}
We explore structural and dynamical behavior of concentrated colloidal suspensions made up by C-shape particles using Brownian dynamics computer simulations and theory.
 In particular, we focus on the entanglement process between nearby particles for almost closed C-shapes with a small opening angle.
Depending on the opening angle and the particle concentration, there is a percolation transition for the cluster of entangled particles
which shows the classical scaling characteristics. 
In a broad density range below the percolation threshold, we find a stretched exponential function for the dynamical decorrelation of the entanglement process.
 Finally, we study a set-up typical in microrheology by dragging a single tagged particle with constant speed through the suspension.
 We measure the cluster connected to and dragged with this tagged particle.
 In agreement with a phenomenological theory, the size of the dragged cluster depends on the dragging direction and increases markedly with the dragging speed.
\end{abstract}


\maketitle

\section{Introduction}

In recent years, the research in the structure and dynamics of colloidal
dispersions has shifted from spherical particles to colloids with an
anisotropic shape \cite{Manoharan_Pine_Science_2003,Glotzer,Dijkstra,our_book,Wittkowski}.
In particular, various forms of hard non-convex particles can be synthesized
by now which are governed by excluded-volume  interaction. The non-convex shape
gives rise to non-trivial close-packed structures and also affects the dynamics
in concentrated suspensions. One of the simplest non-convex shapes is
a hard dumbbell, which has been synthesized
\cite{Johnson2005,Demiroers2010,Gerbode2010}
and was studied quite a lot theoretically
\cite{Vega1992,Marechal_Dijsktra_2008,Wojciechowski1991,Marechal_dumbbells,Dumbbells_Dzubiella,dumbbells_Holm}.
More pronounced non-convexity is realized for colloidal bowls which may penetrate and get stacked \cite{Marechal_bowls}.
While these are still rotational symmetric shapes, more complex
non-convex particles with no rotational symmetry have been considered
including bent rods such as banana-shaped \cite{banana1,banana2}
and boomerang-shaped \cite{boomerang1,Kaiser1,Kaiser2}
particles up to closed rings  or tori \cite{Gabbrielli_2014,Torquato_2012}. Ring-like particles
can get completely entangled with important consequence for their
dynamics and rheological properties.

In this paper we consider colloidal particles with a horseshoe-like shape
 which we will refer to as {\it C-particles}. These
particles are neither rotational symmetric nor convex
and play an important role of linking continuously between slightly bent banana-shapes and full tori. The simplest form of a C-shape is a circular part
where a segment characterized by an opening angle $\alpha$ is cut out from a full circle.
If $\alpha$ vanishes, a ring-like particle results, while for
$\alpha$ close to $2\pi$ we end up with only slightly bent rods. Therefore,
C-particles constitute important interpolating shapes between rods and rings
which can show significant, but not perfect entanglement. It is important to note
that C-shape colloids can be nowadays fabricated at wish
 \cite{Sacanna_Nature_2010,Paulsen_2015} such that our model is realized.

 Here, we explore the structural and dynamical behavior of concentrated colloidal suspensions in three spatial dimensions made up by C-shape particles by using Brownian dynamics computer simulations and theory. In particular, we focus on the entanglement process between nearby particles for almost closed C-shapes with a small
opening angle $\alpha$. By a suitable definition of entanglement, we associate entangled clusters in the suspension and explore their percolation properties \cite{Staufferbook}. Connectivity properties have been studied a lot recently by using concepts of percolation theory, in particular in suspensions of rod-like particles with sticky interactions
\cite{Schilling1,Schilling2,Schilling3,Schilling4,Otten1,Otten2,Nigro,Kyrylyuk}, but
have never systematically been applied to entangled particles.
Depending on the opening angle and the particle concentration,
 we find a percolation transition for the cluster of entangled particles
and identify the percolation threshold. In a broad density range below the percolation threshold,
we find a stretched exponential function $\propto \exp (- \sqrt{t/t_0})$
for the dynamical decorrelation
of the entanglement process between a particle pair where $t_0$ is a characteristic time scale for
disentanglement. This is similar in spirit, but different in detail due to the fluctuations in shape for ring polymers,
to the disentanglement of ring polymers in three dimensions \cite{Turner1,Turner2, Turner3, Threading4}, see also Ref. \cite{Everaers}.

Finally we study a set-up typical in microrheology by dragging a single tagged particles with constant speed through the suspension. We measure the cluster connected to and dragged with this tagged particle. The size of the dragged cluster depends on the dragging direction and increases markedly with the dragging speed. This is due to a dynamical sweeping-up effects mediated by entanglement.
All our predictions can be verified in real-space experiments on colloids by tracking positions and orientations of anisotropic particles \cite{ten_Hagen_Kraft_PRE_2013}.

This paper is organized as follows: the model is described in section II. Equilibrium properties
are discussed in chapter III including both percolation aspects and the dynamics of  disentanglement.
Chapter IV is devoted to the non-equilibrium dragging set-up and we conclude in section V.

\section{Model and methods}

\subsection{Particle model}

\begin{figure}[t]
 \centering
 \includegraphics[width=1.0\linewidth]{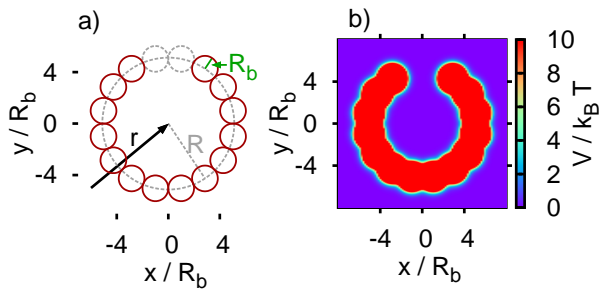}
 \caption{
(a) Sketch of a C-particle of effective radius $R$ as composed of $N_b=14$ hard beads of
 radius  $R_b$. Since $N_m=2$ spheres
are missing to cover the full circle (dashed line), there
is a resulting opening angle $\alpha$. A projection onto the symmetry plane
which contains the full circle is shown.
(b) Softened Yukawa-segment interaction potential energy $V(x,y)$ as seen by a Yukawa bead of another particle,
in the same symmetry plane as in (a) and
 for the same number of beads, with interaction parameters $\kappa R_b=5$ and $V_0/k_BT =1$.}
 \label{fig:model}
\end{figure}

Let us first define our model for C-particles in three dimensions. Ideally we consider a C-particle as composed of $N_b$
spherical hard core beads of radius $R_b$, as shown in Fig. \ref{fig:model}(a). In detail, we start from a full circle
 (dashed line in Fig. \ref{fig:model}(a)) and place in total
$N_b+N_m$ hard spherical beads along the circle such that they are touching. Then, $N_m$ ``missing'' beads are removed
such that $N_b$ beads remain, which gives rise to the characteristic C-shape and introduces an opening angle
\begin{equation}
 \alpha=2 \pi \frac{N_m}{N_b+N_m} .
\end{equation}
We refer to the plane containing all bead centers as the ``symmetry plane'' of the particle.
The resulting radius of the C-particle  (see Fig. \ref{fig:model}(a)) is
\begin{equation}
 R = \epsilon R_b,
\end{equation}
where the aspect ratio $\epsilon$ is given by 
\begin{equation}
\label{eq:epsilon}
 \epsilon=\frac{R}{R_b}=\sqrt{\frac{2}{1-\text{cos}\left(\frac{2\pi}{N_b+N_m}\right)}} .
\end{equation}
In three dimensions, a single C-particle has a physical volume of 
\begin{equation}
 V_C=\frac{4 \pi}{3} R_b^3 N_b 
\end{equation}
and its configuration is fully specified by its central position $\vec{r}$ and a set of three angles $\vec{\varpi}$
describing its orientation in space \cite{GrayGubbins, Wittkowski}.

\begin{figure}[t]
 \centering
 \includegraphics{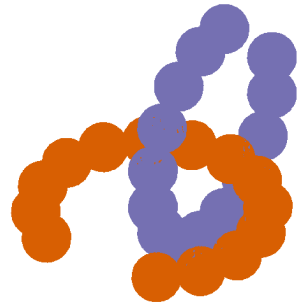}
 \caption{Sketch of two entangled C-particles in three dimensions.}
 \label{fig:entangle}
\end{figure}

We now consider a pair of two C-particles, see Fig. \ref{fig:entangle},
 which can be characterized by their central positions
 ${\vec{r}}_1$ and  ${\vec{r}}_2$ and their orientations ${\vec{\varpi}}_1$ and ${\vec{\varpi}}_2$. 
 Their interaction is steric or of excluded volume-type, i.e.
they are not allowed to overlap. As characteristic for athermal hard-core particles, 
 we assign a vanishing potential energy for a non-overlapping pair and 
a potential energy of infinity for an overlapping pair, the pair potential is
\begin{equation}
 V({\vec{r}}_1 - {\vec{r}}_2 , {\vec{\varpi}}_1, {\vec{\varpi}}_2 ) =  \begin{cases} \infty \text{, if there is overlap} \\
 0 \text{, elsewise}. \end{cases}
\end{equation}

The following criterion for {\it entanglement} of a particle pair is adopted:
for a given configuration of two C-particles, their  missing  spherical beads are formally inserted, i.e.
the C-particle is closed to a full ring such that the opening angle would be vanishing.
Then there are three possibilities; i) the two rings can be moved continuously away from
 each other without crossing any energetic barrier, we refer to this as a {\it disentangled\/} pair, ii) 
the two rings are topologically internested, iii) the two rings physically overlap. By definition, we call
 configurations {\it entangled} if they are not disentangled, i.e.\ if they belong to either case ii) or iii).

\begin{figure}[b]
 \centering
 \includegraphics[width=1.0\linewidth]{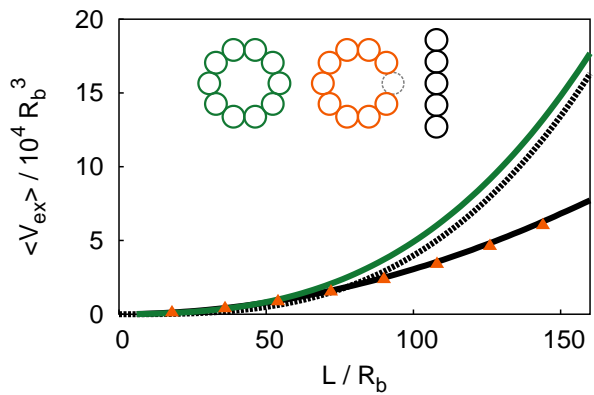}
 \caption{The excluded volume of different objects as a function of the contour length. 
For closed rings (green line), entangled states were not accounted for, 
such that $V_{ex}$ behaves approximately as for hard platelets of the same radius (dashed black line).
 Hard C-particles (here $\alpha=\pi/5$, orange triangles) have approximately the same excluded volume as hard cylindrical rods of the same length (solid black line).}
 \label{fig:excluded}
\end{figure}

As a first immediate result, we address the excluded volume $V_{ex}$ of two hard C-particles.
This quantity measures the degree of steric interactions as embodied in the second virial coefficient
 and is essential, e.g.\ in Onsager's 
theory of the isotropic-nematic transition \cite{Onsager,Lekkerkerker}. The excluded volume $V_{ex}$ can be defined as 
\begin{equation}
\label{eq:Vex}
 V_{ex} = \int \text{d} \vec{r} \left< 1-\exp ( - V(\vec{r} , {\vec{\varpi}}_1, {\vec{\varpi}}_2 ))\right>_{{\vec{\varpi}}_2},
\end{equation}
where the orientational average is given by
\begin{equation}
 < A >_{{\vec{\varpi}}_2} = \frac{1}{8 \pi^2} 
\int \limits_0^{2\pi} \text{d}\chi_2 \int \limits_0^\pi \text{d}\theta_2 \sin{\theta_2} \int \limits_0^{2\pi} \text{d}\phi_2 { A }
\end{equation}
when $\vec{\varpi}_2=(\phi_2,\theta_2,\chi_2)$ is chosen to represent the Euler angles describing the orientation of a rigid particle with non-symmetric shape \cite{GrayGubbins}.
By global rotational symmetry, the definition of $V_{ex}$ in Eq. (\ref{eq:Vex}) does not depend on  ${\vec{\varpi}}_1$. 
We have calculated the excluded volume for different C-shapes using standard Monte-Carlo simulation \cite{Book_Frenkel_Smit}.
Results for $V_{ex}$ for  different particle shapes are presented in Fig. \ref{fig:excluded} 
as a function of the contour length
\begin{equation}
\label{eq:length}
	L=2 \pi R \left( 1 - \frac{\alpha}{2\pi}\right)
\end{equation}
and compared to the straight rod case where \cite{balberg1987} 
\begin{equation}
\left<V_{ex}\right> \approx \pi R_b L^2.
\end{equation}

As can be deduced from Fig. \ref{fig:excluded}, the excluded volume $V_{ex}$ of C-particles basically follows that of rods of the same contour length.
This is qualitatively different if one excludes entangled configurations, as exemplified for closed rings 
for which we have calculated $V_{ex}$ without accounting for entangled configurations, see Fig. \ref{fig:excluded}.
In fact, for $\epsilon \gg 1$, the scaling of $V_{ex}$ in this case is with $L^3$, similar to that of thin platelets,
also shown for comparison in Fig. \ref{fig:excluded}, where $V_{ex}=\pi^2 R_p^3$ with $R_p=R$.
 
Since hard objects are more difficult to simulate within 
 Brownian dynamics simulations \cite{Cichocki_Hinsen,Loewen_PRE_1994,Sciortino,Heyes,Strating}, we softened 
 the interaction between two beads and treat them as harsh Yukawa beads. This has been successfully done for
rod-like particles under various conditions \cite{Kirchhoff_Klein_Loewen_1996,Wensink_JPCM_2012}.
We treat the bead-bead pair interactions $V_{bb}(r)$ as 
\begin{equation}
  V_{bb}(r)=V_0 \frac{2 R_b}{r} \text{exp}(-\kappa(r-2R_b)),
\end{equation}
where the parameters were chosen to be $V_0=V_{bb}(r=2R_b)=k_\text{B} T$ and $\kappa R_b=5$ throughout the simulations. The Yukawa-potential is truncated at $3R_b$. 
The corresponding potential energy landscape felt by a single segment in the neighborhood of a fixed C-particle is shown in Fig. \ref{fig:model}(b). 
The energy quickly increases as the segment approaches the real particle shape such that a mutual crossing in the dynamics is never observed in practice.

\subsection{Brownian Dynamics simulation}

The motion of colloidal C-particles is governed by completely overdamped 
 Brownian dynamics with thermal noise arising from the solvent.
We neglect hydrodynamic interactions between particles, but account for the anisotropic shape of the particles by using
an anisotropic mobility tensor.
Generally, the diffusion tensor of a particle is given by \cite{happel_brenner,de_la_Torre,Doi} 
\begin{equation}
 \mat{D}=k_\text{B} T \mat{H}^{-1},
\end{equation}
where $\mat{H}$ is the grand resistance matrix, which, at low Reynolds number, 
connects forces and torques acting on particle $i$ to its translational and angular velocity.
Here, $k_\text{B} T$ is the thermal energy.
The diffusion tensor  can be written as
\begin{equation}
 \mat{D}=\left(
   \begin{array}{cc}
      \mat{D}_t & \mat{D}^t_c\\
      \mat{D}_c & \mat{D}_r
   \end{array}
\right)
\end{equation}
with the pure translational and rotational submatrices $\mat{D}_t$ and $\mat{D}_r$, and $\mat{D}_c$ 
coupling translational and rotational motion \cite{happel_brenner}.

Hydrodynamically, the C-particles are approximated as oblate ellipsoids of revolution 
with major semi-axes of length $R+R_b$ and a minor semi-axis of length $R_b$ so that
the diffusion tensor reduces to
\begin{equation}
 \mat{D}=\left(
   \begin{array}{cc}
      \mat{D}_t & \mat{0}\\
      \mat{0} & \mat{D}_r
   \end{array}
\right)
\end{equation}
with the analytically known \cite{oberbeck1876, edwardes1893, perrin1934} matrices given in Appendix \ref{app:ellipsoid}, which are diagonal in the body frame of the particle. 
We remark that more sophisticated calculations could be done to match the actual friction coefficients for C-particles better \cite{ten_Hagen_Kraft_PRE_2013,CWagner},
 but we do not expect big changes on the disentangle behavior.

This way, rotational and translational motion decouple \cite{happel_brenner} and the equations of motion are
\begin{equation}
\label{EoMtrans_system}
	\dot{\vec{r}}_i=\beta \mat{R}_i \cdot \mat{D}_t \cdot (\vec{F}_i'+\vec{\xi}_i') 
\end{equation}
and
\begin{equation}
\label{EoMrot_system}
	\vec{\omega}_i=\beta \mat{R}_i \cdot \mat{D}_r \cdot (\vec{T}_i'+\vec{\zeta}_i'),
\end{equation}
where $\beta=(k_\text{B} T)^{-1}$,
$\vec{r}_i$ is the center of mass of particle $i$, $\vec{\omega}_i$ its angular velocity, 
and $\mat{R}_i$ the rotation matrix describing the transformation from the body to the laboratory frame.
 This matrix clearly depends on the orientation $\vec{\varpi}_i$ of particle $i$, i.e. $\mat{R}_i \equiv \mat{R}(\vec{\varpi}_i)$.
Therefore, multiplication with $\mat{R}_i$ transforms a body frame vector to the laboratory frame, while its inverse does the reverse transformation,
 e.g. $\vec{F}_i = \mat{R}_i \cdot \vec{F}_i'$ and $\vec{F}_i' = \mat{R}_i^{-1} \cdot \vec{F}_i$.
Here, primed vectors belong to the body frame, and vectors without a prime symbol to the laboratory frame.

Moreover, the interaction force $\vec{F}_i$ on particle $i$ due to the other particles is given by
\begin{equation}
 \vec{F}_i=\sum_{b \in i} \vec{F}_{b}^{(i)},
\end{equation}
where $\vec{F}_{b}^{(i)}$ is the force on bead $b$ of particle $i$ due to the Yukawa-interaction with all other particles' beads
and the sum is over all beads $b$ of particle $i$.
The corresponding torque $\vec{T}_i$ is determined by 
\begin{equation}
 \vec{T}_i=\sum_{b\in i} (\vec{r}_b^{(i)}-\vec{r}_i) \times \vec{F}_b^{(i)},
\end{equation}
where $\vec{r}_b^{(i)}$ denotes the bead center.
The stochastic Brownian force $\vec{\xi}_i$ is Gaussian-distributed with zero mean and second moment
\begin{equation}
 \left< \xi_{i\alpha}'(t) \xi_{j\beta}'(t') \right> =  
 k_\text{B} T H_{\alpha \alpha} \delta_{ij}\delta_{\alpha \beta} \delta(t-t') 
\end{equation}
in the body frame, where the axes are the principal axes of the particle, with $\alpha$ and $\beta$ denoting Cartesian components.
In close analogy, the Gaussian Brownian torque $\vec{\zeta}_i$ has zero mean and second moment
 \begin{equation}
 \left< \zeta_{i\alpha}'(t) \zeta_{j\beta}'(t') \right> =  
  k_\text{B} T H_{(\alpha+3) (\alpha+3)} \delta_{ij} \delta_{\alpha \beta} \delta(t-t') 
\end{equation}
in the body frame.

\begin{figure}[t]
\centering
\includegraphics{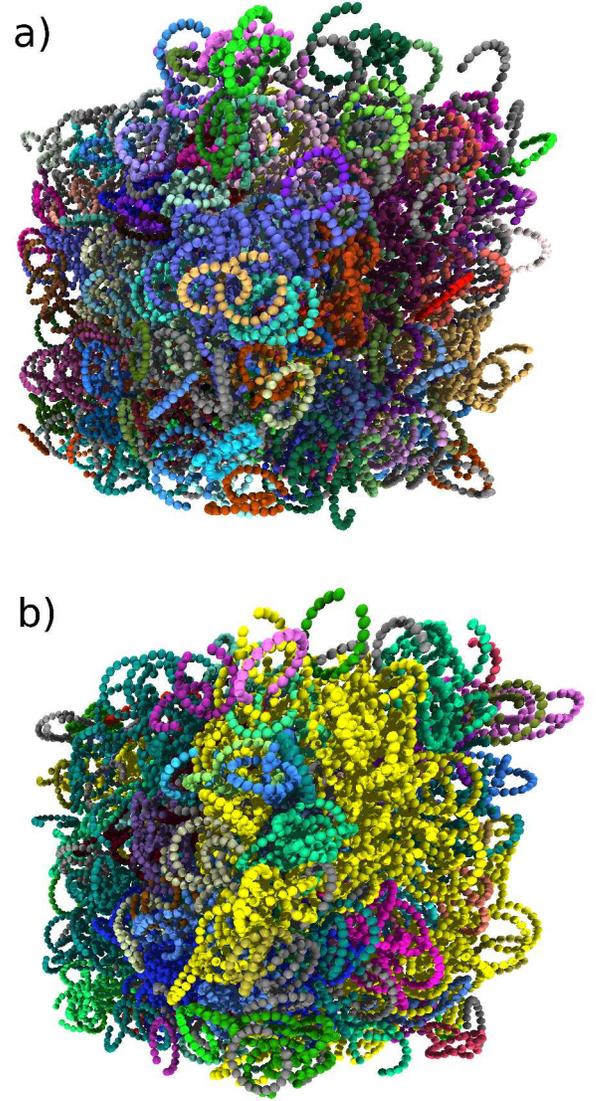}
\caption{Simulation snapshots of a system with $N=1000$ particles and shape parameters $\epsilon=6.37$ and $\alpha=\pi/5$
below the percolation threshold at density $nR^3=0.274$ (a) and percolated at density $nR^3=0.342$ (b). 
Particles belonging to the same cluster are shown in the same color. 
Particles with no entanglement are colored gray, the giant component in (b) is colored yellow.
}
 \label{fig:snapshots}
\end{figure}

Eqs. \ref{EoMtrans_system} and \ref{EoMrot_system} are numerically solved using the standard Euler-forward scheme 
for stochastic differential equations \cite{Ermak_1975_JCP}
\begin{equation}
\label{eq:EoMtrans_discrete}
	\vec{r}_i(t+\Delta t) - \vec{r}_i(t) = \mat{R}_i(t) \cdot 
\left( \beta \Delta {t} \mat{D}_t \cdot \vec{F}_i'(t) + {\vec{\sigma}}_i' \right)
\end{equation}
and (going into the body frame)
\begin{equation}
\label{eq:EoMrot_body_discrete}
\vec{\omega}_i'(t) \Delta t = \beta \Delta{t} \mat{D}_r \cdot \vec{T}_i'(t) +{\vec{\psi}}_i'
\end{equation}
with time step $\Delta t$. 
Here, the Brownian forces $\vec{\xi}_i$ and torques $\vec{\zeta}_i$ have been integrated over time and thus replaced by Brownian translational and rotational moves,
$\vec{\sigma}_i$ and $\vec{\psi}_i$, which are again Gaussian with zero mean and second moment
\begin{equation}
 \left< \sigma_{i\alpha}' \sigma_{i\beta}' \right> =  
 2 \delta_{\alpha \beta} D_{t, \alpha \alpha} \Delta t 
\end{equation}
and
 \begin{equation}
 \left< \psi_{i\alpha}' \psi_{i\beta}' \right> =  
 2 \delta_{\alpha \beta} D_{r, \alpha \alpha} \Delta t ,
\end{equation}
respectively.
Moreover, we used Beard and Schlick's method for bias-free rotational moves \cite{biasfreerotation}.

 In the simulation, the coupled equations of motion (\ref{EoMtrans_system}) and (\ref{EoMrot_system}) for $N$ C-particles are solved.
The time step was made adaptive to limit particle displacement in one single time step to below $ 0.05 R_b$, with a maximal time step $\Delta t \leq 10^{-4} \tau_\text{B}$,
where the Brownian time is defined as 
\begin{equation}
 \tau_\text{B}=R_b^2/D_{11}.
\end{equation}
The orientations of the particles were stored as quaternions, from which the corresponding rotation matrices $\mat{R}_i$ were computed and
which allowed for fast updates of the orientation.

When not noted otherwise, our results stem from simulations of $N=1000$ particles with an aspect ratio $\epsilon < 10$,
see typical system snapshots in Fig. \ref{fig:snapshots}. We
started with random orientations on equally distributed grid points and  then  relaxed
the system  for a time $10 \tau_\text{B}$, and gathered statistics
 until $40 \tau_\text{B}$. The cubic system was treated with periodic boundary conditions in all three directions, 
where the box length $\ell$ is determined by the  prescribed volume fraction 
\begin{equation}
 \Phi =  N V_C / \ell^3
\end{equation}
 of the C-particles. The entanglement criterion as developed earlier was 
numerically implemented with a piercing technique outlined in Appendix B. For non-vanishing opening angles,
$\alpha$, it was carefully checked that the 
equilibrated system does not exhibit spontaneous nematic ordering but stays  orientationally disordered.

\section{C-particles in equilibrium}

\subsection{Number of entanglements and percolation}

\begin{figure}[t]
 \centering
 \includegraphics[width=1.0\linewidth]{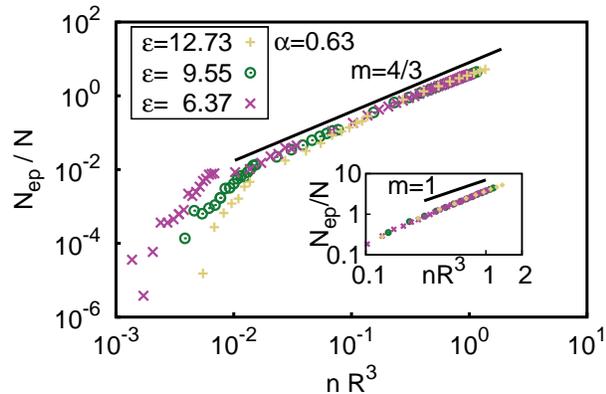}
 \caption{The number of entangled pairs per particle as a function of $nR^3$ for fixed opening angle $\alpha=\pi/5$. 
Apart from very low densities, the graphs collapse. 
The inset shows that the scaling becomes linear for high densities.}
 \label{fig:cp_reduced}
\end{figure}

We have calculated the number of entangled pairs per particle, $N_{ep}/N$, using
 the entanglement criterion introduced above. 
The data in Fig. \ref{fig:cp_reduced} are shown as a function of $nR^3$ since this is the natural 
representation for infinitely thin C-particles, i.e. if the size ratio $\epsilon \to \infty$, 
so that the particles radius $R$ is the only length scale left.
The double-logarithmic plot  reveals that the number of entanglements scales as $n R^3$ for high densities.
 For intermediate densities, we find a long crossover 
which looks to scale as $n^{4/3} R^4$ for one to two decades of $n R^3$.
The high-density linear scaling can be qualitatively explained with the following argument:
Entanglement of one particle can only occur with other particles in a volume $\propto R^3$ around the particle.
Since the physical volume is full of particles at high densities, the probability 
for such particles scales as $nR^3$ in this case. 
Whether the behavior for intermediate density is a real scaling or just a crossover is an interesting question which we leave for future studies.

\begin{figure*}[t]
 \centering
 \includegraphics[width=1.0\linewidth]{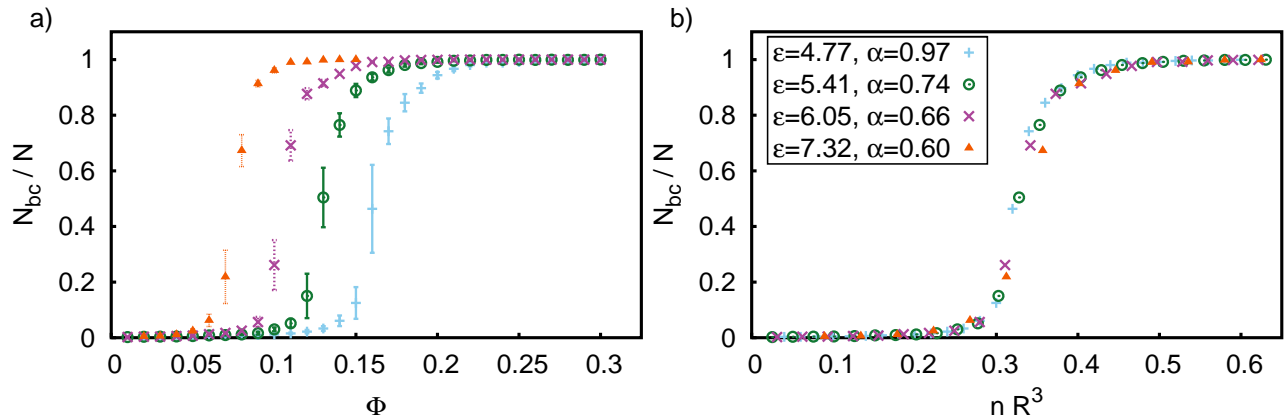}
 \caption{The fraction of particles in the biggest cluster as a function of the volume fraction occupied by C-particles (a) and as a function of the number density times the particle radius cubed (b), for different aspect ratios as given in the legend. When the particles radius is increased, the critical volume fraction decreases. All simulations in these plots contained $N=1000$ particles.}
 \label{fig:bc_reduced}
\end{figure*}

We assign two particles to be in the same entanglement cluster if there is a chain of entangled particles 
connecting them. This criterion is symmetric, transitive, and by definition reflexive. Thus, a 
well-defined sorting of particles is established that divides the system into a discrete set of clusters. As for an example, 
see the simulation snapshots in Fig. \ref{fig:snapshots} 
where different clusters are shown in different colors.
Near the system percolation transition, the biggest cluster of the system involves a finite fraction $N_{bc}/N$ of 
particles in the thermodynamic limit $N \to \infty$. Data for the biggest cluster are presented in Fig. \ref{fig:bc_reduced}
as a function of the volume fraction $\Phi$ (a) and as a function of the reduced density $nR^3$ (b). For aspect ratios $\epsilon\gg 1$, i.e. for slender C-particles, there should only be a 
dependence on the opening angle $\alpha$. Indeed, Fig. \ref{fig:bc_reduced}(b) shows that there is a good scaling on a single Master curve
and that both dependencies on $\epsilon$ and  $\alpha$ are weak. There is a sharp jump from almost vanishing values
for $N_{bc}/N$ towards a finite value at about $nR^3\approx 0.31$ for all $\epsilon$ and  $\alpha$ which
reveals that there is a percolation transition in the system.

\begin{figure}[b]
 \centering
 \includegraphics[width=1.0\linewidth]{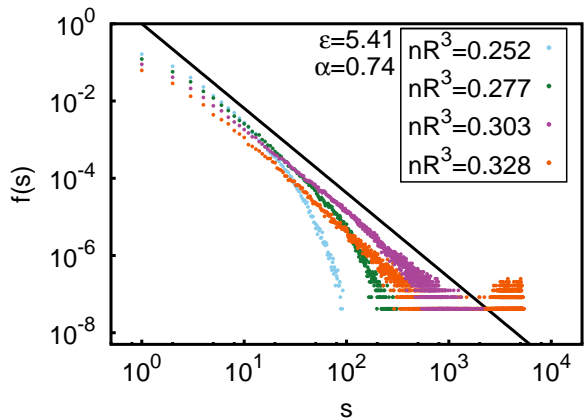}
 \caption{Distribution of cluster sizes for particles with aspect ratio $\epsilon=5.41$ for different reduced densities $nR^3$.
The data were obtained for a large system with $N=8000$ C-particles. At the critical density, the distribution follows a power law with universal exponent $\tau=-2.189(2)$ \cite{lorenz1998}. The corresponding slope is indicated by the solid line.
 Below the percolation threshold, the distribution is exponentially cut-off; above, there is a peak at large sizes, representing the giant component.}
 \label{fig:bc_chain_distri}
\end{figure}

\begin{figure}[b]
 \centering
 \includegraphics[width=1.0\linewidth]{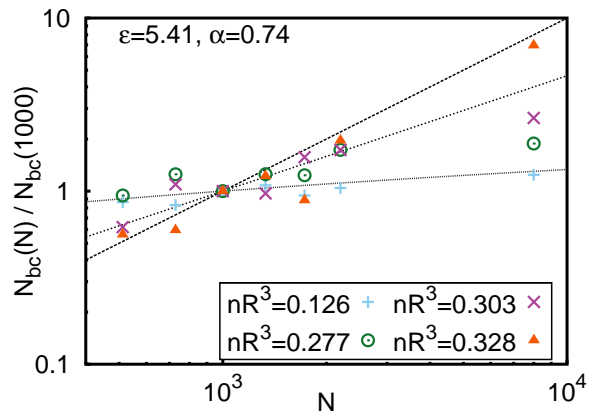}
 \caption{Finite-size scaling of the biggest cluster size $N_{bc}$, 
normalized with its value for a system size of $N=1000$,
with respect to the system size $N$:
 unpercolated ($\propto \text{log}N)$, critical ($\propto N^{2/3}$), and percolated ($\propto N$) systems \cite{newman_networks}. The lines give the expected system size behaviors. The system parameters are $\epsilon=5.41$ and $\alpha=0.74$.
The percolation transition occurs at
$nR^3 \approx 0.31$.
}
 \label{fig:bc_finitesize}
\end{figure}

Figs. \ref{fig:bc_chain_distri} and \ref{fig:bc_finitesize} demonstrate that standard percolation theory \cite{newman_networks, albert_networks} applies 
for this transition. The theory predicts that there is a critical percolation point characterized by 
 universal critical exponents where the biggest cluster fraction is getting non-zero in the thermodynamic limit.

We have checked that two general properties of classical percolation do also apply for our system.
First, we have plotted the cluster size distribution function $f(s)$ with $s$ denoting a size of an entangled cluster
in Fig. \ref{fig:bc_chain_distri}. Below the critical point, the distribution of cluster sizes should follow the law
\begin{equation}
 f(s)\propto s^{-\tau} \text{exp}\left(-\frac{s}{s_\text{co}}\right)
\end{equation}
with the Fisher exponent $\tau=2.189(2)$ for undirected percolation in  three spatial dimensions \cite{lorenz1998}. 
At the critical point, the cut-off size $s_\text{co}$ diverges and $f(s)$ follows a power law with universal exponent $\tau=-2.189(2)$ \cite{lorenz1998},
shown as a reference line in Fig. \ref{fig:bc_chain_distri}. 
Above the critical point, an exponential decay is seen again, but also a peak at large $s$ exists, 
the so-called giant component. This is indeed confirmed by the data shown in Fig. \ref{fig:bc_chain_distri}.

Second, we have checked finite system size corrections near the percolation transition. Classical percolation 
theory \cite{newman_networks} predicts that $N_{bc}$ behaves as  $\propto \text{log}N$
below the percolation transition, $\propto N^{2/3}$ at the percolation transition and 
$\propto N$ above the percolation transitions. The data shown in Fig. \ref{fig:bc_finitesize} indeed confirm these three predictions.

\subsection{Long-time diffusion and dynamics of disentanglement}

\begin{figure}[b]
 \centering
 \includegraphics[width=1.0\linewidth]{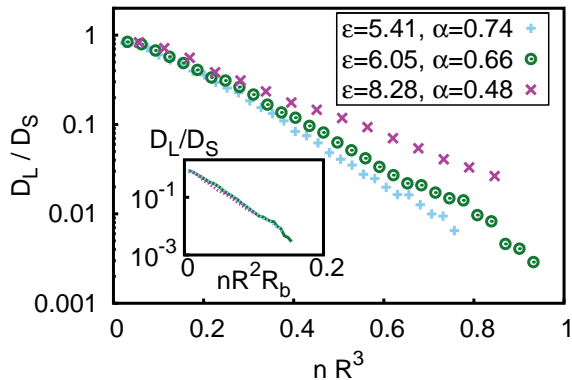}
 \caption{The long-time diffusion constant $D_{L}$ decreases exponentially with increasing density. Larger C-particles feature a stronger decrease.
	  The inset shows that data collapse for $\log_{10}(D_L / D_S)$ as a function of $n R^2 R_b$.}
 \label{fig:difflong}
\end{figure}

We now address the long-time dynamics as embodied in the mean-square displacement of the center-of-mass coordinate
defined as
\begin{equation}
 \Delta (t) = \left< \Big( {\vec r_i}(t) - {\vec r_i}(0) \Big)^2 \right>.
\end{equation}
For short times, the Brownian dynamics is diffusive by construction and the orientationally averaged
short-time diffusion constant can be extracted as
\begin{equation}
 D_S = \lim_{ t \to 0} \Delta (t)/ 6t.
\end{equation}
For long times $t \to \infty$,  the mean-square displacement $\Delta (t)$ is again diffusive,
 such that we define as usual a long-time self-diffusion coefficient 
\begin{equation}
 D_L = \lim_{ t \to \infty} \Delta (t)/ 6t .
\end{equation}
In general, at finite density, $D_L$ is smaller than its short-time counterpart $D_S$
as the dynamics is hindered by the presence of neighboring C-particles and $D_L$ decreases with increasing
density. Numerically, we extract $D_L$ from our simulation data for the mean-square displacement in a finite time window by 
extrapolating the relaxation in $\Delta (t)$ towards long-time diffusion algebraically \cite{lps1993} as 
\begin{equation}
 \Delta(t)=6 D_L t + \frac{A t}{\tau_0+t},
\end{equation}
where the amplitude $A$ and the time scale $\tau_0$ are fit parameters to the simulation data.
Fig. \ref{fig:difflong} shows the density dependence of $D_L/D_S$. 
The inset reveals that a simple exponential dependence on density rescaled with $R^2 R_b$ such that
\begin{equation}
\label{diff_exp}
  D_L/D_S = \exp ( -\lambda n R^2 R_b)
\end{equation}
 is a  good fit for the long-time self-diffusion coefficient over several decades,
at least for parameter combinations considered here. 
 This is interesting as convex Brownian hard spheres \cite{Szamel,2dLinde} and hard spherocylinders \cite{Loewen_PRE_1994} 
exhibit an almost linear dependence of $D_L/D_S$ on density up to the freezing density. Here, non-convexity
and the resulting entanglements slow down the dynamics more drastically, being close to  glass formation.
In fact, in binary hard sphere mixtures \cite{Wiesner,Williams} also an exponential increase  of 
the characteristic relaxation time in the density was found for densities not too close to the glass transition.
This corresponds to an exponential decrease of $D_L$ as one can extract a typical relaxation time $\tau_L$ as
\begin{equation}
 \tau_L = R^2/D_L,
\end{equation}
which is the characteristic time to diffuse over the size of a C-particle.

We grouped particles with respect to how many entanglements they have on average during the whole simulation length,
 and then looked at the mean-squared displacements in these groups.
For unpercolated systems, we find that the more entanglements a particle has, the more slowly it diffuses.
When the system is fully percolated ($N_{bc}\approx N$),
 there is no more any difference and all particles diffuse with the same constant.
This indicates that not only the number of entanglements pins down C-particles and thus hinders diffusion, 
but that the connectivity is a further obstacle for diffusion.

\begin{figure*}[t]
 \centering
 \includegraphics[width=1.0\linewidth]{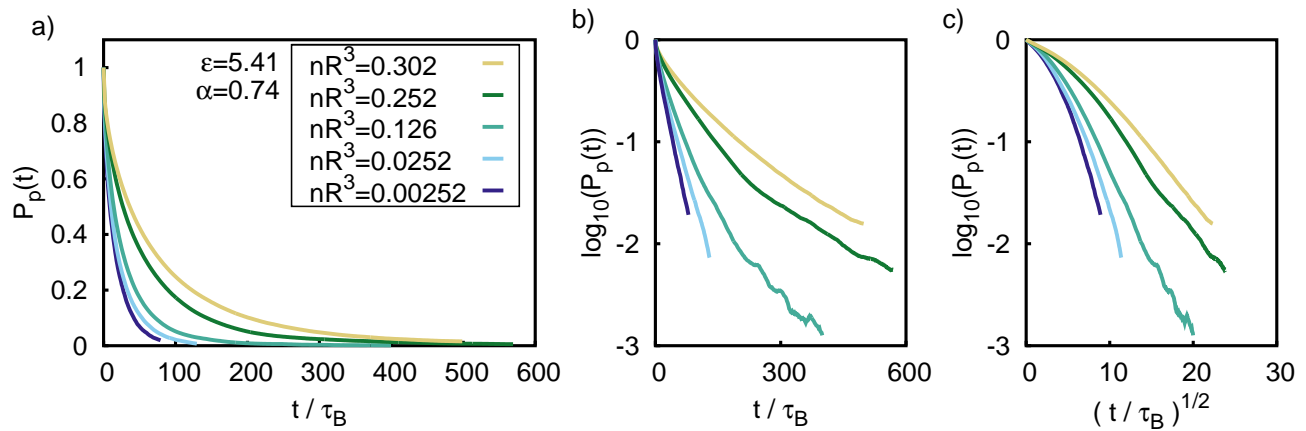}
 \caption{(a) The entanglement correlation function for C-particles with aspect ratio $\epsilon=5.41$ and opening angle $\alpha=0.74$ for different densities $n$ as a function of time. (b,c) The logarithm of the entanglement correlation function as a function of the time (b) and as function of the square root of time (c), for the same data as (a). For low densities, there is a simple exponential decay (straight lines in (b), 
for higher densities and long times a stretched exponential with exponent $\beta=1/2$ is found (straight lines in (c)).
}
 \label{fig:persistence_overview}
\end{figure*}

\begin{figure}[b]
 \centering
 \includegraphics[width=1.0\linewidth]{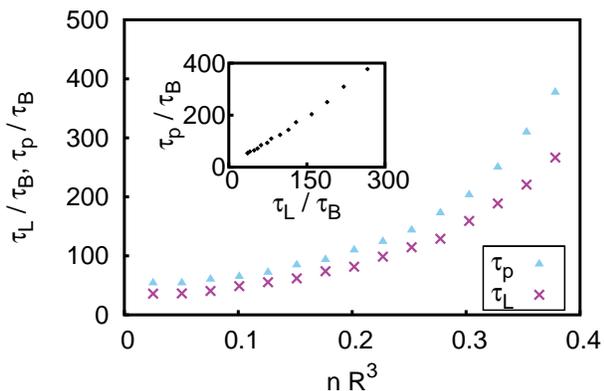}
 \caption{ Disentanglement time scale $\tau_p$ and long-time diffusional time scale $\tau_L$ as functions of the density
 for C-particles with aspect ratio $\epsilon=5.41$ and opening angle $\alpha=0.74$. 
The inset shows the proportionality between the two time scales, using the same data as the main plot.}
 \label{fig:taup}
\end{figure}

We now define a dynamical autocorrelation function describing the disentanglement process.
First, we define the observable
 \begin{equation}
 p(i,j;t)=\begin{cases}
	    1 \text{, if } (i, j) \text{ are entangled at time } t \\
	    0 \text{, elsewise}
          \end{cases}
\end{equation}
as an entanglement order parameter and consider its time autocorrelation function, given by
\begin{equation}
 P_p(t)=\frac{\left< p(i,j;t_0+t)p(i,j;t_0)\right>}{\left< p(i,j;t_0) p(i,j;t_0)\right>},
\end{equation}
where the brackets denote averages over particle pairs $(i,j)$ and initial times $t_0$.
Here, the term in the denominator can be further simplified, since $p^2(i,j;t)=p(i,j;t)$.
 Thereby, $P_p(t=0)=1$ is normalized at zero time. Conversely, for long times,
the entanglement order parameters are statistically independent such that
 \begin{equation}
\label{eq:longtime-P}
 \begin{aligned}
  \lim_{t \to \infty}P_p(t)
  &=\frac{\left< p(i,j;t_0+t)\right> \left< p(i,j;t_0)\right>}
  {\left< p(i,j;t_0))\right>}\\
  &=\left< p(i,j;t_0)\right>\\
  &=\frac{2 N_{ep}}{N(N-1)}.
  \end{aligned}
 \end{equation}
 As $N_{ep} \propto N$, the long-time limit of $P_p(t)$ vanishes in the thermodynamic limit.   

\begin{figure*}[t]
 \centering
 \includegraphics[width=1.0\linewidth]{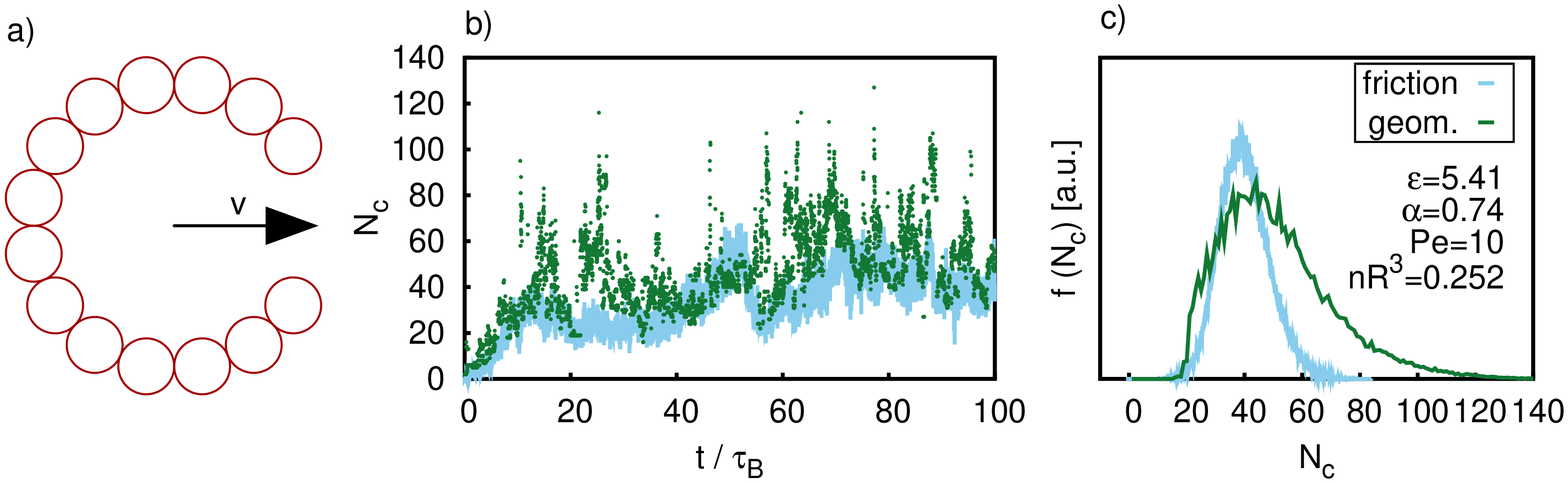}
 \caption{
(a) Forward drag scenario,
(b) pulled cluster size as determined structurally and via friction,
(c) distribution function of the  pulled cluster size measured structurally and via friction, for the same data as (b).
}
 \label{fig:friction_paper}
\end{figure*}

Fig. \ref{fig:persistence_overview}(a) shows data for the entanglement autocorrelation function $P_p(t)$ on a linear scale
which exhibit a decay that is significantly slowed down for increasing densities. In fact, a semi-logarithmic plot
(see Figs. \ref{fig:persistence_overview}(b,c)) shows that either a normal exponential or a stretched exponential decay in time occurs.
The former occurs for small densities as expected, the latter happens for densities which are below but not too
close to the percolation threshold. Here, the stretched exponential form 
\begin{equation}
 P_p(t) = \exp \left( - (t/\tau_p)^\beta \right)
\end{equation}
is a good fit to the data which contains the time scale $\tau_p$ as a typical relaxation time of the disentanglement process.
As Fig. \ref{fig:persistence_overview}(c) shows, the value for the exponent $\beta$ is close to 1/2.
A similar stretched exponential relaxation scenario is obtained in the context of glass formation where it is 
better known as a Kohlrausch-Williams-Watt law and results from cage relaxation. Computer 
simulations \cite{Angelani,Rabani} as well as experiments and mode-coupling theory \cite{Goetze_book} give indications that 
the exponent $\beta=1/2$ can be realized in the context of glasses as well.

In fact, the typical relaxation time $\tau_p$, defined via $P_p(\tau_p)=0.1$, is closely correlated to $\tau_L$ which sets the time for long-time 
self-diffusion. This is documented in Fig. \ref{fig:taup}, where we plot both time scales as a function of the density in the main plot
and $\tau_p$ as a function of $\tau_L$ in the inset.

At this stage, we remark that the threading of ring-polymers is an analogue to the entanglement of C-particles. 
These threadings, i.e. one ring penetrating another one, 
and their influence on the dynamics of a ring-polymer system were explored by simulation in a recent series of papers \cite{Turner1, Turner2, Turner3}.  
Indeed, a significant slow-down of heavily threaded systems and
a stretched exponential decay in the time-correlation function of an observable measuring contiguity, i.e. closeness, of two ring-polymers
was found where the exponent approaches $\beta=1/2$ for sufficiently large rings \cite{Turner3}.
Therefore, regarding C-particles and their entanglement as a ``topologically driven glass'' as introduced in Refs. \cite{Turner1,Turner2,Turner3}
looks to have merit and might be a fruitful subject for future examinations. 

\section{Microrheology of C-particle suspensions}

We now address a non-equilibrium aspect of C-particles which is closely connected to entanglement.
A single ``tagged''  particle is dragged with prescribed velocity  $v$,
 constituting a set-up of constant-velocity microrheology \cite{puertas_microrheology}. 

\subsection{Simulation scheme}

In our constant-velocity microrheology \cite{puertas_microrheology} simulations, we  suspend
 the Brownian dynamics move scheme for one ``tagged'' particle. Instead, its center-of-mass coordinate
 is moved each time step by 
$v \Delta t \uvec{u} $, where $\uvec{u}$ is a constant unit vector giving the drag direction, $v$ is the dragging speed
and the particle's orientation is kept fixed.

First, we consider a situation in which the particle is dragged towards the opening of the C, see Fig. \ref{fig:friction_paper}(a). 
As a second case, the probing particle is pulled in exactly the opposite direction. The former favors 
absorbing new particles into the pulled cluster, while the latter leads to particles slipping off the
 tagged particle's cluster.
The simulations were started from equilibrated starting positions. The data were then averaged over $15$ different runs
 each lasting 100 $\tau_\text{B}$.

\subsection{Forward drag}

\begin{figure*}[t]
 \centering
 \includegraphics[width=1.0\linewidth]{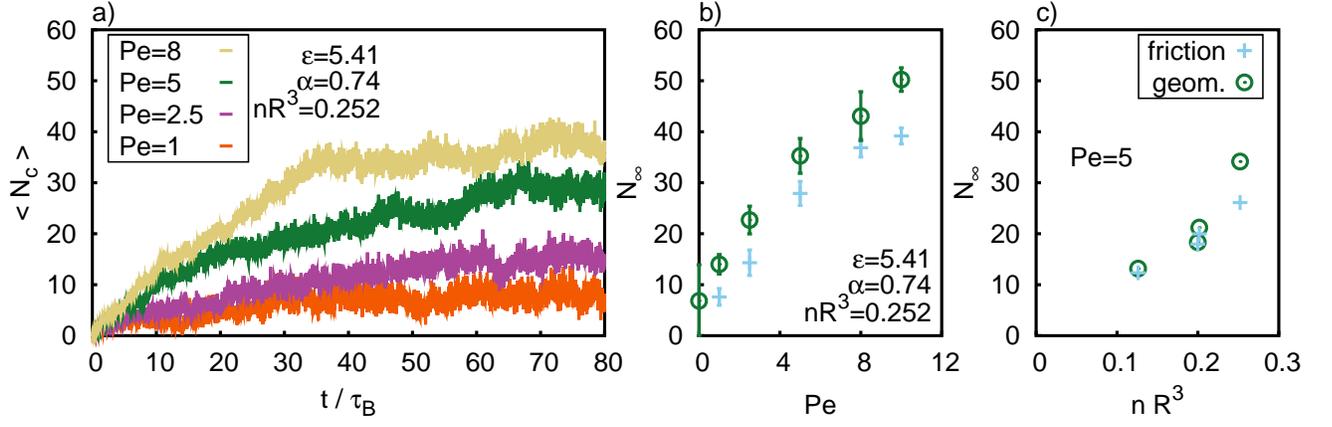}
 \caption{
(a) The time evolution of the pulled cluster size for different drag speeds, all starting from the same equilibrium state, each averaged over $15$ runs. 
(b, c) The mean steady-state size of the pulled cluster, when varying the P\'eclet number (b) and the density  $n R^3$ (c), measured structurally and via friction.
}
 \label{fig:pulling_paper}
\end{figure*}

Here, the dragging direction $\uvec{u}$ shows from the particle's center towards the opening of the ring, see Fig. \ref{fig:friction_paper}(a).
During the run, the force on the dragged particle, $\vec{F}_p$, was recorded. 
While it has no effect on the particle's velocity in our setup, it 
would have to be compensated in an experimental situation to achieve this constant speed.
Therefore it may also serve to define a frictional cluster size $N_c^\text{fric}$ as follows.
We average the projection of this force onto the drag direction, $ <\vec{F}_p \cdot \uvec{u}>$, and 
postulate that this is equal to the hydrodynamic drag force of $N_c^\text{fric}$ single uncorrelated particles,
i.e. we define $N_c^\text{fric}$ via the relation 
\begin{equation}
	- <\vec{F}_p \cdot \uvec{u}> = N_\text{c}^\text{fric} H_\uvec{u} v ,
\end{equation}
with $H_\uvec{u}$ being the friction coefficient along the $\uvec{u}$-axis of the body frame of a single particle.

Simultaneously, we measured $N_\text{c}$ using the structural entanglement cluster criterion as defined earlier.
An example is presented  in Fig. \ref{fig:friction_paper}(b). The structurally-defined cluster size has much higher fluctuations
than the frictional one, as is documented also by the corresponding distribution function in the steady state, see Fig. \ref{fig:friction_paper}(c),
which is much sharper for
$N_c^\text{fric}$ than for $N_\text{c}$. The high overshoots in $N_\text{c}$ correspond to situations where 
blocks of particles are just about to leave the cluster such that they are still counted structurally without having any
contribution to the frictional force. 
Fig. \ref{fig:pulling_paper}(a) shows that the  cluster attached to
 the tagged particle - as averaged over different initial conditions - grows with time,
 but approaches a finite long-time limit indicative of a steady state.

We shall now model this dynamical process by using phenomenological arguments,
 following similar ideas as in earlier work \cite{Carnevale,Kolb,Hansen,Wensink,Cremer}.
Basically, the cluster dynamics is fixed by two processes, one leading to growth and the other to loss
of cluster size. For the {\it growth process\/} we assume that the pulled C-particle collects all 
particles which it hits with its opening cross section of area $A_c=\alpha R R_b$. Per time unit
it will cover a volume $A_c v$ such that on average  $n A_c v$ particles per time unit will be swept 
up or gathered into the moving C-particle. Each of this particle is connected with other C-particles
 according to the equilibrium mean size $\left<s\right>_p$ of entangled clusters. Hence, we obtain for the growth process
the following rate of particle getting per time unit attached to the moving cluster
\begin{equation}
 \dot{N}_\text{c}^+ = nv \left< s \right>_p A_c.
\end{equation}

Conversely, for the {\it loss process\/}, we assume that a particle only leaves the cluster 
when its opening shows in the direction of $\uvec{u}$. Assuming single particle rotational Brownian dynamics 
and an initial perpendicular configuration, the typical time to reach the opening scales 
with $\tau_R = (2\pi - \alpha)^{2} / D_{66}$, 
where $D_{66}$ is the diffusion constant for rotations around the particle's normal vector $\uvec{n}$.
 When a particle strips away, it also removes all 
entangled particles behind it from the pulled cluster, leading to a proportionality on $\left<s\right>_p$
for the loss rate. As the loss process can result from any individual particle in the cluster, there is an additional proportionality to the actual cluster size ${N}_\text{c}$ such that we get the total loss rate
\begin{equation}
 \dot{N}_\text{c}^- \propto \left<s\right>_p N_c /\tau_r.
\end{equation}

Combining the two equations, we obtain
\begin{equation}
 \dot{N}_\text{c} = \dot{N}_\text{c}^+ + \dot{N}_\text{c}^- =  nv\left<s\right>_p A_c  - C_l \left<s\right>_p D_{66} N_c
\end{equation}
where $C_l$ is a numerical constant. With $D_{66}\approx D_{11} R^{-2}$ (see Appendix \ref{app:ellipsoid}) and $A_c \propto R_b R$, we arrive at
\begin{equation}
\label{eq:pull_ode}
 \dot{N}_\text{c}=
  C_g nv\left<s\right>_p R R_b- C_l \left<s\right>_p {D}_{11} R^{-2} N_c,
\end{equation}
where $C_g$ is another numerical constant. Solving this ordinary differential equation with the initial condition 
$N_c(t=0)=\left<s\right>_p$ gives 
\begin{equation}
 N_c(t)= \left<s\right>_p \text{exp}\left(-\frac{t}{\tau_\text{gr}}\right)+N_\infty\left(1-\text{exp}\left(-\frac{t}{\tau_\text{gr}}\right)\right)
\end{equation}
with the \emph{growth time}
\begin{equation}
 \tau_\text{gr}=\frac{R^2}{C_l\left<s\right>_p {D}_{11}}
\end{equation}
and the steady-state size of the dragged cluster
\begin{equation}
 N_\infty=\lim_{t \to \infty}N_c=\frac{C_g}{C_l} n R^3  \text{Pe}, 
\end{equation}
where we introduced the P\'eclet number 
\begin{equation}
 \text{Pe}=v R_b / {D}_{11}.
\end{equation}

Two scaling predictions of our theory are tested in Fig. \ref{fig:pulling_paper}. In Fig. \ref{fig:pulling_paper}(b),
simulation data for the steady-state cluster size $N_\infty$ are shown versus P\'eclet number $\text{Pe}$ for 
fixed density and there is indeed linear scaling of $N_\infty$  versus $\text{Pe}$
over a broad range of P\'eclet numbers as predicted by the theory. It is only for 
high P\'eclet numbers that deviations from the linear dependence point to the relevance of
steric exclusion effects inside the dragged C-particle.
Similarly, $N_\infty$ scales roughly linearly in 
density $nR^3$ for fixed P\'eclet number $\text{Pe}$, see  Fig. \ref{fig:pulling_paper}(c), thereby confirming 
the theoretical scaling prediction.

\subsection{Backward drag}

\begin{figure}[b]
 \centering
 \includegraphics[width=1.0\linewidth]{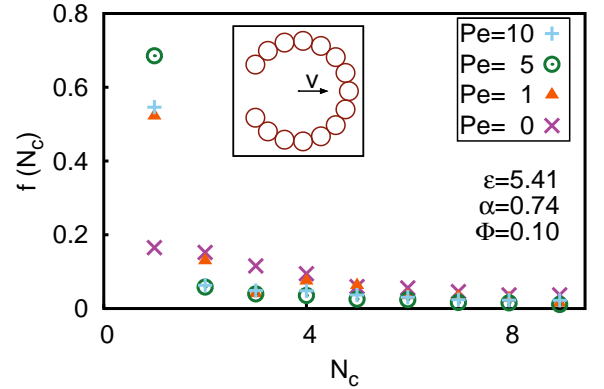}
 \caption{Distribution function $f(N_c)$ of the pulled cluster size,
 measured structurally by the entanglement cluster criterion.
The inset shows  the backward drag situation schematically.}
 \label{fig:pulling_loss}
\end{figure}

Let us finally consider dragging a particle away from its opening, where we expect a shrinkage of the dragged cluster. 
As shown in Fig. \ref{fig:pulling_loss}, the distribution of the pulled cluster size 
indeed indicates shrinking with increasing P\'eclet number. In fact, the pulled particle moves mostly alone, 
no entanglement persists for a longer time. Thus, $\left< N_{c} \right>$ is getting smaller than its equilibrium counterpart 
$\left<s\right>_p$. For extremely large $\text{Pe}$, the mean increases again because the pulled particle 
encounters more particles which are then part of its cluster for a short time. 
Hence, interestingly there is a minimum of $\left<N_{c}\right>$ as a function of $\text{Pe}$.

\section{Conclusions}

In conclusion, we have explored the structure and dynamics of entangling colloidal horseshoe-like 
particles (referred to as C-particles) by using Brownian dynamics computer simulations. 
There is a percolation transition of mutually 
entangled structures which shows a similar signature as percolation in other three-dimensional particulate systems.
In the dense regime, the disentanglement dynamics was found to be governed by a stretched exponential 
demonstrating the dynamical arrest caused by entanglement. Finally, we have found a profound 
impact of the entangling process on 
the microrheological behavior where a single particle is dragged through the suspension carrying a whole wake
of entangled particles which follow the dragged one. We have identified scaling laws describing this
dragged-induced  accumulation process.

Our results are in principles verifiable in colloidal suspensions of non-convex particles. Moreover, 
we expect that most of the qualitative features do also hold for granulates
 (such as paper-clips etc.) under microgravity \cite{Stannarius}.
Future work should  address the entanglement dynamics in other systems involving particles with 
other non-convex shapes than considered here. This could maybe include L-particles with a sharp cusp \cite{Bechinger1,Bechinger2} and
semiflexible curved polymers (see e.g. Ref. \cite{Yoshikawa}). It would be interesting
 to check the universality of the disentanglement dynamics. The presence of a sharp cusp in the particle shape
is expected to increase entanglement effects significantly.

\begin{acknowledgments}
The authors thank Matthieu Marechal, Liesbeth M.C. Janssen and Peet Cremer for helpful comments.
Support from the 
European Research Council under the European Union's Seventh Framework Programme, ERC Grant Agreement No. 267499,
 is acknowledged.
\end{acknowledgments}

\appendix

\section{Diffusion tensor of an ellipsoid of revolution}
\label{app:ellipsoid}
The following diffusion coefficients for oblate ellipsoids of revolution \cite{oberbeck1876, edwardes1893, perrin1934} were used:
\begin{widetext}
 \begin{equation}
 16 \pi \eta a \mat{D}_t=\left(
   \begin{array}{ccc}
     \dfrac{(3\xi^2-2) S - 1}{\xi^2-1} & . & .\\
    . &  \dfrac{(3\xi^2-2) S - 1}{\xi^2-1} & .\\
    . & . & 2 \dfrac{(\xi^2-2) S + 1}{\xi^2-1}
   \end{array}
\right)
\end{equation}
and
\begin{equation}
 \frac{16\pi}{3} \eta a^3 \mat{D}_r=\left(
   \begin{array}{ccc}
     \begin{array}{cccccc}
    \dfrac{(\xi^2-2)S+1}{\xi^4-1} & . & .\\
    . & \dfrac{(\xi^2-2)S+1}{\xi^4-1} & .\\
    . & . & \dfrac{S-\xi^{-2}}{\xi^2-1}
   \end{array}
   \end{array}
\right),
\end{equation}
where $\xi=(R+R_b)/R_b=\epsilon+1$ is the ratio of the lengths of the long semi-axes and the short semi-axis, and
\begin{equation}
 S=\frac{1}{\sqrt{\xi^2-1}}\text{arctan} \left( \sqrt{\xi^2-1} \right).
\end{equation}
Here, we used the right-handed coordinate system $(\uvec{\ell},\uvec{m},\uvec{n})$ where $\uvec{n}$ is perpendicular to the symmetry plane, $\uvec{\ell}$ is parallel to the line between the particle's center and the center of its opening, and $\uvec{m}=\uvec{n}\times \uvec{\ell}$.
\end{widetext}

\section{Entanglement criterion}
\label{sec:entanglement_criterion}

\begin{figure}[t]
 \centering
 \includegraphics[width=1.0\linewidth]{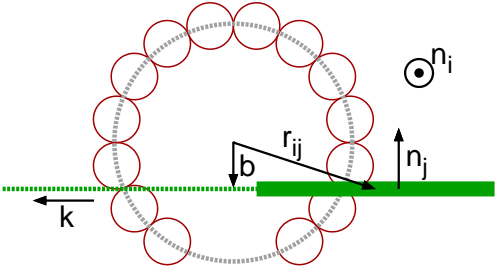}
 \caption{Vectors involved in the entanglement criterion. The green rectangle represents the ring of C-particle $j$ which is oriented perpendicular to C-particle $i$ here.}
 \label{fig:entangle_schematic}
\end{figure}

In the following, we present a simple criterion of entanglement, for which we just need the distance vector $\vec{r}_{ij}:=\vec{r}_i-\vec{r}_j$ between the centers of mass of two particles $i$ and $j$, and the unit vectors $\uvec{n}_i$ and $\uvec{n}_j$ normal to their symmetry planes, see Fig. \ref{fig:entangle_schematic} for an exemplary situation.

The geometrical idea from which we start is the following: 
Consider the two discs described by the beads of the particles. If there is an entanglement, the first disc's rim pierces the other disc in exactly one point, and vice versa. If they do not pierce or pierce in two points, there is no entanglement. Thus, the number of piercing points is crucial for entanglement.

All points common to both discs of the particles fulfill
\begin{equation}
\label{pierce:ebene}
	(\vec{r}-\vec{r}_i) \cdot \uvec{n}_i = 0 = (\vec{r}-\vec{r}_j) \cdot \uvec{n}_j .
\end{equation}
Thus, these points are described by $\vec{r}=\vec{r}_i+\vec{b}+\lambda \vec{k}$, where $\vec{k}= \uvec{n}_i \times \uvec{n}_j$
and $\vec{b}$ is a vector pointing from the center of mass of particle $i$ to the common line.
 It follows that
\begin{equation}
	\vec{b} \cdot \uvec{n}_i = 0 = (\vec{b}+\vec{r}_{ij}) \cdot \uvec{n}_j .
\end{equation}
We further set
\begin{equation}
  \label{pierce:bk}
	\vec{b} \cdot \vec{k}=0 ,
\end{equation}
which simplifies Eq. (\ref{pierce:lambda}), and get
\begin{equation}
	\vec{b}= \frac{\uvec{n}_j\cdot \vec{r}_{ij}}{k^2} 
	[ (\uvec{n}_i \cdot \uvec{n}_j) \, \uvec{n}_i - \uvec{n}_j ]
\end{equation}
 
For entangled $i$ and $j$, two points $\vec{r}_{1,2}$ are found
which lie in both planes, i.e. $\vec{r}_{1,2}=\vec{r}_i+\vec{b}+\lambda_{1,2}\vec{k}$,
and on the line of circle $i$, i.e.
\begin{equation}
\label{pierce:fixed}
	(\vec{r}_{1,2}-\vec{r}_i)^2 = R^2,
\end{equation}
 and
of which one lies on disk $j$, while the other one does not, i.e.
\begin{align}
\label{pierce:2}
\begin{aligned}
	(\vec{r}_{1}-\vec{r}_j)^2 &> R^2,
	\\(\vec{r}_{2}-\vec{r}_j)^2 &< R^2.
\end{aligned}
\end{align}
To check for entanglement, we thus look if such two points exist, i.e. if a pair of $(\lambda_1, \lambda_2)$ with these properties exists.

Combining the ansatz for $\vec{r}$ and Eqs. \ref{pierce:bk} and \ref{pierce:fixed} leads to
\begin{equation}
  \label{pierce:lambda}
	R^2=(\vec{b}+\lambda_{1,2} \vec{k})^2= b^2 + \lambda^2_{1,2} k^2,
\end{equation}
which gives $\lambda_{1,2} = \pm \lambda$ with $k^2\lambda^2=(R^2-b^2).$ 
 If $R^2<b^2,$ there are no points fulfilling the given conditions.

Inserting
\begin{equation}
\begin{aligned}
(\vec{r}_{1,2}-\vec{r}_j)^2&=(\vec{b} \pm \lambda\vec{k}+\vec{r}_{ij})^2
\\&=R^2+r_{ij}^2 \pm 2\lambda\vec{k}\cdot\vec{r}_{ij}+2\vec{b}\cdot\vec{r}_{ij},
\end{aligned}
\end{equation}
into Eq. (\ref{pierce:2}) leads to
\begin{align}
\begin{aligned}
	r_{ij}^2 + 2\lambda\vec{k}\cdot\vec{r}_{ij}+2\vec{b}\cdot\vec{r}_{ij} &> 0,
	\\ r_{ij}^2 - 2\lambda\vec{k}\cdot\vec{r}_{ij}+2\vec{b}\cdot\vec{r}_{ij} &< 0.
\end{aligned}
\end{align}
These two inequalities are only simultaneously solvable if
\begin{equation}
 (R^2-b^2) (\vec{k}\cdot\vec{r}_{ij})^2 > 
  k^2 \left(\frac{r_{ij}^2}{2}  + \vec{b}\cdot\vec{r}_{ij}\right)^2.
\end{equation}
For any two given particles, this criterion can be easily checked numerically.

\bibliographystyle{apsrev}
\bibliography{c_particles}

\begin{thebibliography}{81}
\expandafter\ifx\csname natexlab\endcsname\relax\def\natexlab#1{#1}\fi
\expandafter\ifx\csname bibnamefont\endcsname\relax
  \def\bibnamefont#1{#1}\fi
\expandafter\ifx\csname bibfnamefont\endcsname\relax
  \def\bibfnamefont#1{#1}\fi
\expandafter\ifx\csname citenamefont\endcsname\relax
  \def\citenamefont#1{#1}\fi
\expandafter\ifx\csname url\endcsname\relax
  \def\url#1{\texttt{#1}}\fi
\expandafter\ifx\csname urlprefix\endcsname\relax\def\urlprefix{URL }\fi
\providecommand{\bibinfo}[2]{#2}
\providecommand{\eprint}[2][]{\url{#2}}

\bibitem[{\citenamefont{Manoharan et~al.}(2003)\citenamefont{Manoharan,
  Elsesser, and Pine}}]{Manoharan_Pine_Science_2003}
\bibinfo{author}{\bibfnamefont{V.~N.} \bibnamefont{Manoharan}},
  \bibinfo{author}{\bibfnamefont{M.~T.} \bibnamefont{Elsesser}},
  \bibnamefont{and} \bibinfo{author}{\bibfnamefont{D.~J.} \bibnamefont{Pine}},
  \bibinfo{journal}{Science} \textbf{\bibinfo{volume}{301}},
  \bibinfo{pages}{483} (\bibinfo{year}{2003}).

\bibitem[{\citenamefont{Glotzer and Solomon}(2007)}]{Glotzer}
\bibinfo{author}{\bibfnamefont{S.~C.} \bibnamefont{Glotzer}} \bibnamefont{and}
  \bibinfo{author}{\bibfnamefont{M.~J.} \bibnamefont{Solomon}},
  \bibinfo{journal}{Nat. Mater.} \textbf{\bibinfo{volume}{6}},
  \bibinfo{pages}{557} (\bibinfo{year}{2007}).

\bibitem[{\citenamefont{Dijkstra}(2014)}]{Dijkstra}
\bibinfo{author}{\bibfnamefont{M.}~\bibnamefont{Dijkstra}},
  \bibinfo{journal}{Adv. Chem. Phys.} \textbf{\bibinfo{volume}{156}},
  \bibinfo{pages}{35} (\bibinfo{year}{2014}).

\bibitem[{\citenamefont{Ivlev et~al.}(2012)\citenamefont{Ivlev, L{\"o}wen,
  Morfill, and Royall}}]{our_book}
\bibinfo{author}{\bibfnamefont{A.}~\bibnamefont{Ivlev}},
  \bibinfo{author}{\bibfnamefont{H.}~\bibnamefont{L{\"o}wen}},
  \bibinfo{author}{\bibfnamefont{G.}~\bibnamefont{Morfill}}, \bibnamefont{and}
  \bibinfo{author}{\bibfnamefont{C.~P.} \bibnamefont{Royall}},
  \emph{\bibinfo{title}{Complex Plasmas and Colloidal Dispersions:
  Particle-Resolved Studies of Classical Liquids and Solids}}
  (\bibinfo{publisher}{World Scientific}, \bibinfo{year}{2012}).

\bibitem[{\citenamefont{Wittkowski and L{\"o}wen}(2011)}]{Wittkowski}
\bibinfo{author}{\bibfnamefont{R.}~\bibnamefont{Wittkowski}} \bibnamefont{and}
  \bibinfo{author}{\bibfnamefont{H.}~\bibnamefont{L{\"o}wen}},
  \bibinfo{journal}{Mol. Phys.} \textbf{\bibinfo{volume}{109}},
  \bibinfo{pages}{2935} (\bibinfo{year}{2011}).

\bibitem[{\citenamefont{Johnson et~al.}(2005)\citenamefont{Johnson, van Kats,
  and van Blaaderen}}]{Johnson2005}
\bibinfo{author}{\bibfnamefont{P.~M.} \bibnamefont{Johnson}},
  \bibinfo{author}{\bibfnamefont{C.~M.} \bibnamefont{van Kats}},
  \bibnamefont{and} \bibinfo{author}{\bibfnamefont{A.}~\bibnamefont{van
  Blaaderen}}, \bibinfo{journal}{Langmuir} \textbf{\bibinfo{volume}{21}},
  \bibinfo{pages}{11510} (\bibinfo{year}{2005}).

\bibitem[{\citenamefont{Demir\"ors et~al.}(2010)\citenamefont{Demir\"ors,
  Johnson, van Kats, van Blaaderen, and Imhof}}]{Demiroers2010}
\bibinfo{author}{\bibfnamefont{A.~F.} \bibnamefont{Demir\"ors}},
  \bibinfo{author}{\bibfnamefont{P.~M.} \bibnamefont{Johnson}},
  \bibinfo{author}{\bibfnamefont{C.~M.} \bibnamefont{van Kats}},
  \bibinfo{author}{\bibfnamefont{A.}~\bibnamefont{van Blaaderen}},
  \bibnamefont{and} \bibinfo{author}{\bibfnamefont{A.}~\bibnamefont{Imhof}},
  \bibinfo{journal}{Langmuir} \textbf{\bibinfo{volume}{26}},
  \bibinfo{pages}{14466} (\bibinfo{year}{2010}).

\bibitem[{\citenamefont{Gerbode et~al.}(2010)\citenamefont{Gerbode, Agarwal,
  Ong, Liddell, Escobedo, and Cohen}}]{Gerbode2010}
\bibinfo{author}{\bibfnamefont{S.~J.} \bibnamefont{Gerbode}},
  \bibinfo{author}{\bibfnamefont{U.}~\bibnamefont{Agarwal}},
  \bibinfo{author}{\bibfnamefont{D.~C.} \bibnamefont{Ong}},
  \bibinfo{author}{\bibfnamefont{C.~M.} \bibnamefont{Liddell}},
  \bibinfo{author}{\bibfnamefont{F.}~\bibnamefont{Escobedo}}, \bibnamefont{and}
  \bibinfo{author}{\bibfnamefont{I.}~\bibnamefont{Cohen}},
  \bibinfo{journal}{Phys. Rev. Lett.} \textbf{\bibinfo{volume}{105}},
  \bibinfo{pages}{078301} (\bibinfo{year}{2010}).

\bibitem[{\citenamefont{Vega et~al.}(1992)\citenamefont{Vega, Paras, and
  Monson}}]{Vega1992}
\bibinfo{author}{\bibfnamefont{C.}~\bibnamefont{Vega}},
  \bibinfo{author}{\bibfnamefont{E.~P.~A.} \bibnamefont{Paras}},
  \bibnamefont{and} \bibinfo{author}{\bibfnamefont{P.~A.}
  \bibnamefont{Monson}}, \bibinfo{journal}{J. Chem. Phys.}
  \textbf{\bibinfo{volume}{97}}, \bibinfo{pages}{8543} (\bibinfo{year}{1992}).

\bibitem[{\citenamefont{Marechal and Dijkstra}(2008)}]{Marechal_Dijsktra_2008}
\bibinfo{author}{\bibfnamefont{M.}~\bibnamefont{Marechal}} \bibnamefont{and}
  \bibinfo{author}{\bibfnamefont{M.}~\bibnamefont{Dijkstra}},
  \bibinfo{journal}{Phys. Rev. E} \textbf{\bibinfo{volume}{77}},
  \bibinfo{pages}{061405} (\bibinfo{year}{2008}).

\bibitem[{\citenamefont{Wojciechowski et~al.}(1991)\citenamefont{Wojciechowski,
  Frenkel, and Bra{\'n}ka}}]{Wojciechowski1991}
\bibinfo{author}{\bibfnamefont{K.~W.} \bibnamefont{Wojciechowski}},
  \bibinfo{author}{\bibfnamefont{D.}~\bibnamefont{Frenkel}}, \bibnamefont{and}
  \bibinfo{author}{\bibfnamefont{A.}~\bibnamefont{Bra{\'n}ka}},
  \bibinfo{journal}{Phys. Rev. Lett.} \textbf{\bibinfo{volume}{66}},
  \bibinfo{pages}{3168} (\bibinfo{year}{1991}).

\bibitem[{\citenamefont{Marechal et~al.}(2011)\citenamefont{Marechal, Goetzke,
  H{\"a}rtel, and L{\"o}wen}}]{Marechal_dumbbells}
\bibinfo{author}{\bibfnamefont{M.}~\bibnamefont{Marechal}},
  \bibinfo{author}{\bibfnamefont{H.~H.} \bibnamefont{Goetzke}},
  \bibinfo{author}{\bibfnamefont{A.}~\bibnamefont{H{\"a}rtel}},
  \bibnamefont{and}
  \bibinfo{author}{\bibfnamefont{H.}~\bibnamefont{L{\"o}wen}},
  \bibinfo{journal}{J. Chem. Phys.} \textbf{\bibinfo{volume}{135}},
  \bibinfo{pages}{234510} (\bibinfo{year}{2011}).

\bibitem[{\citenamefont{Chu et~al.}(2015)\citenamefont{Chu, Heptner, Lu,
  Siebenb\"urger, Lindner, Dzubiella, and Ballauff}}]{Dumbbells_Dzubiella}
\bibinfo{author}{\bibfnamefont{F.}~\bibnamefont{Chu}},
  \bibinfo{author}{\bibfnamefont{N.}~\bibnamefont{Heptner}},
  \bibinfo{author}{\bibfnamefont{Y.}~\bibnamefont{Lu}},
  \bibinfo{author}{\bibfnamefont{M.}~\bibnamefont{Siebenb\"urger}},
  \bibinfo{author}{\bibfnamefont{P.}~\bibnamefont{Lindner}},
  \bibinfo{author}{\bibfnamefont{J.}~\bibnamefont{Dzubiella}},
  \bibnamefont{and} \bibinfo{author}{\bibfnamefont{M.}~\bibnamefont{Ballauff}},
  \bibinfo{journal}{Langmuir} \textbf{\bibinfo{volume}{31}},
  \bibinfo{pages}{5992} (\bibinfo{year}{2015}).

\bibitem[{\citenamefont{Fischer et~al.}(2015)\citenamefont{Fischer, Peter,
  Holm, and de~Graaf}}]{dumbbells_Holm}
\bibinfo{author}{\bibfnamefont{L.~P.} \bibnamefont{Fischer}},
  \bibinfo{author}{\bibfnamefont{T.}~\bibnamefont{Peter}},
  \bibinfo{author}{\bibfnamefont{C.}~\bibnamefont{Holm}}, \bibnamefont{and}
  \bibinfo{author}{\bibfnamefont{J.}~\bibnamefont{de~Graaf}},
  \bibinfo{journal}{J. Chem. Phys.} \textbf{\bibinfo{volume}{143}},
  \bibinfo{pages}{084107} (\bibinfo{year}{2015}).

\bibitem[{\citenamefont{Marechal et~al.}(2010)\citenamefont{Marechal,
  Kortschot, Demir\"ors, Imhof, and Dijkstra}}]{Marechal_bowls}
\bibinfo{author}{\bibfnamefont{M.}~\bibnamefont{Marechal}},
  \bibinfo{author}{\bibfnamefont{R.~J.} \bibnamefont{Kortschot}},
  \bibinfo{author}{\bibfnamefont{A.~F.} \bibnamefont{Demir\"ors}},
  \bibinfo{author}{\bibfnamefont{A.}~\bibnamefont{Imhof}}, \bibnamefont{and}
  \bibinfo{author}{\bibfnamefont{M.}~\bibnamefont{Dijkstra}},
  \bibinfo{journal}{Nano Lett.} \textbf{\bibinfo{volume}{10}},
  \bibinfo{pages}{1907} (\bibinfo{year}{2010}).

\bibitem[{\citenamefont{Memmer}(2002)}]{banana1}
\bibinfo{author}{\bibfnamefont{R.}~\bibnamefont{Memmer}},
  \bibinfo{journal}{Liq. Cryst.} \textbf{\bibinfo{volume}{29}},
  \bibinfo{pages}{483} (\bibinfo{year}{2002}).

\bibitem[{\citenamefont{{Mart{\'\i}nez-Gonz{\'a}lez}
  et~al.}(2012)\citenamefont{{Mart{\'\i}nez-Gonz{\'a}lez}, Varga, Gurin, and
  Quintana-H}}]{banana2}
\bibinfo{author}{\bibfnamefont{J.}~\bibnamefont{{Mart{\'\i}nez-Gonz{\'a}lez}}},
  \bibinfo{author}{\bibfnamefont{S.}~\bibnamefont{Varga}},
  \bibinfo{author}{\bibfnamefont{P.}~\bibnamefont{Gurin}}, \bibnamefont{and}
  \bibinfo{author}{\bibfnamefont{J.}~\bibnamefont{Quintana-H}},
  \bibinfo{journal}{EPL} \textbf{\bibinfo{volume}{97}}, \bibinfo{pages}{26004}
  (\bibinfo{year}{2012}).

\bibitem[{\citenamefont{Chakrabarty et~al.}(2013)\citenamefont{Chakrabarty,
  Konya, Wang, Selinger, Sun, and Wei}}]{boomerang1}
\bibinfo{author}{\bibfnamefont{A.}~\bibnamefont{Chakrabarty}},
  \bibinfo{author}{\bibfnamefont{A.}~\bibnamefont{Konya}},
  \bibinfo{author}{\bibfnamefont{F.}~\bibnamefont{Wang}},
  \bibinfo{author}{\bibfnamefont{J.~V.} \bibnamefont{Selinger}},
  \bibinfo{author}{\bibfnamefont{K.}~\bibnamefont{Sun}}, \bibnamefont{and}
  \bibinfo{author}{\bibfnamefont{Q.-H.} \bibnamefont{Wei}},
  \bibinfo{journal}{Phys. Rev. Lett.} \textbf{\bibinfo{volume}{111}},
  \bibinfo{pages}{160603} (\bibinfo{year}{2013}).

\bibitem[{\citenamefont{Kaiser et~al.}(2014)\citenamefont{Kaiser, Peshkov,
  Sokolov, ten Hagen, L{\"o}wen, and Aranson}}]{Kaiser1}
\bibinfo{author}{\bibfnamefont{A.}~\bibnamefont{Kaiser}},
  \bibinfo{author}{\bibfnamefont{A.}~\bibnamefont{Peshkov}},
  \bibinfo{author}{\bibfnamefont{A.}~\bibnamefont{Sokolov}},
  \bibinfo{author}{\bibfnamefont{B.}~\bibnamefont{ten Hagen}},
  \bibinfo{author}{\bibfnamefont{H.}~\bibnamefont{L{\"o}wen}},
  \bibnamefont{and} \bibinfo{author}{\bibfnamefont{I.~S.}
  \bibnamefont{Aranson}}, \bibinfo{journal}{Phys. Rev. Lett.}
  \textbf{\bibinfo{volume}{112}}, \bibinfo{pages}{158101}
  (\bibinfo{year}{2014}).

\bibitem[{\citenamefont{Kaiser et~al.}(2015)\citenamefont{Kaiser, Sokolov,
  Aranson, and L\"owen}}]{Kaiser2}
\bibinfo{author}{\bibfnamefont{A.}~\bibnamefont{Kaiser}},
  \bibinfo{author}{\bibfnamefont{A.}~\bibnamefont{Sokolov}},
  \bibinfo{author}{\bibfnamefont{I.}~\bibnamefont{Aranson}}, \bibnamefont{and}
  \bibinfo{author}{\bibfnamefont{H.}~\bibnamefont{L\"owen}},
  \bibinfo{journal}{Eur. Phys. J. Special Topics}
  \textbf{\bibinfo{volume}{224}}, \bibinfo{pages}{1275} (\bibinfo{year}{2015}).

\bibitem[{\citenamefont{Gabbrielli et~al.}(2014)\citenamefont{Gabbrielli, Jiao,
  and Torquato}}]{Gabbrielli_2014}
\bibinfo{author}{\bibfnamefont{R.}~\bibnamefont{Gabbrielli}},
  \bibinfo{author}{\bibfnamefont{Y.}~\bibnamefont{Jiao}}, \bibnamefont{and}
  \bibinfo{author}{\bibfnamefont{S.}~\bibnamefont{Torquato}},
  \bibinfo{journal}{Phys. Rev. E} \textbf{\bibinfo{volume}{89}},
  \bibinfo{pages}{022133} (\bibinfo{year}{2014}).

\bibitem[{\citenamefont{Torquato and Jiao}(2012)}]{Torquato_2012}
\bibinfo{author}{\bibfnamefont{S.}~\bibnamefont{Torquato}} \bibnamefont{and}
  \bibinfo{author}{\bibfnamefont{Y.}~\bibnamefont{Jiao}},
  \bibinfo{journal}{Phys. Rev. E} \textbf{\bibinfo{volume}{86}},
  \bibinfo{pages}{011102} (\bibinfo{year}{2012}).

\bibitem[{\citenamefont{Sacanna et~al.}(2010)\citenamefont{Sacanna, Irvine,
  Chaikin, and Pine}}]{Sacanna_Nature_2010}
\bibinfo{author}{\bibfnamefont{S.}~\bibnamefont{Sacanna}},
  \bibinfo{author}{\bibfnamefont{W.~T.~M.} \bibnamefont{Irvine}},
  \bibinfo{author}{\bibfnamefont{P.~M.} \bibnamefont{Chaikin}},
  \bibnamefont{and} \bibinfo{author}{\bibfnamefont{D.~J.} \bibnamefont{Pine}},
  \bibinfo{journal}{Nature} \textbf{\bibinfo{volume}{464}},
  \bibinfo{pages}{575} (\bibinfo{year}{2010}).

\bibitem[{\citenamefont{Paulsen et~al.}(2015)\citenamefont{Paulsen, Di~Carlo,
  and Chung}}]{Paulsen_2015}
\bibinfo{author}{\bibfnamefont{K.~S.} \bibnamefont{Paulsen}},
  \bibinfo{author}{\bibfnamefont{D.}~\bibnamefont{Di~Carlo}}, \bibnamefont{and}
  \bibinfo{author}{\bibfnamefont{A.~J.} \bibnamefont{Chung}},
  \bibinfo{journal}{Nat. Commun.} \textbf{\bibinfo{volume}{6}}
  (\bibinfo{year}{2015}).

\bibitem[{\citenamefont{Stauffer and Aharony}(1994)}]{Staufferbook}
\bibinfo{author}{\bibfnamefont{D.}~\bibnamefont{Stauffer}} \bibnamefont{and}
  \bibinfo{author}{\bibfnamefont{A.}~\bibnamefont{Aharony}},
  \emph{\bibinfo{title}{Introduction to percolation theory}}
  (\bibinfo{publisher}{CRC press}, \bibinfo{year}{1994}).

\bibitem[{\citenamefont{Meyer et~al.}(2015)\citenamefont{Meyer, Van~der Schoot,
  and Schilling}}]{Schilling1}
\bibinfo{author}{\bibfnamefont{H.}~\bibnamefont{Meyer}},
  \bibinfo{author}{\bibfnamefont{P.}~\bibnamefont{Van~der Schoot}},
  \bibnamefont{and}
  \bibinfo{author}{\bibfnamefont{T.}~\bibnamefont{Schilling}},
  \bibinfo{journal}{J. Chem. Phys.} \textbf{\bibinfo{volume}{143}},
  \bibinfo{pages}{044901} (\bibinfo{year}{2015}).

\bibitem[{\citenamefont{Mathew et~al.}(2012)\citenamefont{Mathew, Schilling,
  and Oettel}}]{Schilling2}
\bibinfo{author}{\bibfnamefont{M.}~\bibnamefont{Mathew}},
  \bibinfo{author}{\bibfnamefont{T.}~\bibnamefont{Schilling}},
  \bibnamefont{and} \bibinfo{author}{\bibfnamefont{M.}~\bibnamefont{Oettel}},
  \bibinfo{journal}{Phys. Rev. E} \textbf{\bibinfo{volume}{85}},
  \bibinfo{pages}{061407} (\bibinfo{year}{2012}).

\bibitem[{\citenamefont{Schilling et~al.}(2007)\citenamefont{Schilling,
  Jungblut, and Miller}}]{Schilling3}
\bibinfo{author}{\bibfnamefont{T.}~\bibnamefont{Schilling}},
  \bibinfo{author}{\bibfnamefont{S.}~\bibnamefont{Jungblut}}, \bibnamefont{and}
  \bibinfo{author}{\bibfnamefont{M.~A.} \bibnamefont{Miller}},
  \bibinfo{journal}{Phys. Rev. Lett.} \textbf{\bibinfo{volume}{98}},
  \bibinfo{pages}{108303} (\bibinfo{year}{2007}).

\bibitem[{\citenamefont{Nigro et~al.}(2013{\natexlab{a}})\citenamefont{Nigro,
  Grimaldi, Miller, Ryser, and Schilling}}]{Schilling4}
\bibinfo{author}{\bibfnamefont{B.}~\bibnamefont{Nigro}},
  \bibinfo{author}{\bibfnamefont{C.}~\bibnamefont{Grimaldi}},
  \bibinfo{author}{\bibfnamefont{M.~A.} \bibnamefont{Miller}},
  \bibinfo{author}{\bibfnamefont{P.}~\bibnamefont{Ryser}}, \bibnamefont{and}
  \bibinfo{author}{\bibfnamefont{T.}~\bibnamefont{Schilling}},
  \bibinfo{journal}{Phys. Rev. E} \textbf{\bibinfo{volume}{88}},
  \bibinfo{pages}{042140} (\bibinfo{year}{2013}{\natexlab{a}}).

\bibitem[{\citenamefont{Otten and van~der Schoot}(2012)}]{Otten1}
\bibinfo{author}{\bibfnamefont{R.~H.~J.} \bibnamefont{Otten}} \bibnamefont{and}
  \bibinfo{author}{\bibfnamefont{P.}~\bibnamefont{van~der Schoot}},
  \bibinfo{journal}{Phys. Rev. Lett.} \textbf{\bibinfo{volume}{108}},
  \bibinfo{pages}{088301} (\bibinfo{year}{2012}).

\bibitem[{\citenamefont{Otten and van~der Schoot}(2011)}]{Otten2}
\bibinfo{author}{\bibfnamefont{R.~H.~J.} \bibnamefont{Otten}} \bibnamefont{and}
  \bibinfo{author}{\bibfnamefont{P.}~\bibnamefont{van~der Schoot}},
  \bibinfo{journal}{J. Chem. Phys.} \textbf{\bibinfo{volume}{134}},
  \bibinfo{pages}{094902} (\bibinfo{year}{2011}).

\bibitem[{\citenamefont{Nigro et~al.}(2013{\natexlab{b}})\citenamefont{Nigro,
  Grimaldi, Ryser, Chatterjee, and Van Der~Schoot}}]{Nigro}
\bibinfo{author}{\bibfnamefont{B.}~\bibnamefont{Nigro}},
  \bibinfo{author}{\bibfnamefont{C.}~\bibnamefont{Grimaldi}},
  \bibinfo{author}{\bibfnamefont{P.}~\bibnamefont{Ryser}},
  \bibinfo{author}{\bibfnamefont{A.~P.} \bibnamefont{Chatterjee}},
  \bibnamefont{and} \bibinfo{author}{\bibfnamefont{P.}~\bibnamefont{Van
  Der~Schoot}}, \bibinfo{journal}{Phys. Rev. Lett.}
  \textbf{\bibinfo{volume}{110}}, \bibinfo{pages}{015701}
  (\bibinfo{year}{2013}{\natexlab{b}}).

\bibitem[{\citenamefont{Kyrylyuk and van~der Schoot}(2008)}]{Kyrylyuk}
\bibinfo{author}{\bibfnamefont{A.~V.} \bibnamefont{Kyrylyuk}} \bibnamefont{and}
  \bibinfo{author}{\bibfnamefont{P.}~\bibnamefont{van~der Schoot}},
  \bibinfo{journal}{Proc. Natl. Acad. Sci.} \textbf{\bibinfo{volume}{105}},
  \bibinfo{pages}{8221} (\bibinfo{year}{2008}).

\bibitem[{\citenamefont{Michieletto
  et~al.}(2014{\natexlab{a}})\citenamefont{Michieletto, Marenduzzo, Orlandini,
  Alexander, and Turner}}]{Turner1}
\bibinfo{author}{\bibfnamefont{D.}~\bibnamefont{Michieletto}},
  \bibinfo{author}{\bibfnamefont{D.}~\bibnamefont{Marenduzzo}},
  \bibinfo{author}{\bibfnamefont{E.}~\bibnamefont{Orlandini}},
  \bibinfo{author}{\bibfnamefont{G.~P.} \bibnamefont{Alexander}},
  \bibnamefont{and} \bibinfo{author}{\bibfnamefont{M.~S.}
  \bibnamefont{Turner}}, \bibinfo{journal}{ACS Macro Lett.}
  \textbf{\bibinfo{volume}{3}}, \bibinfo{pages}{255}
  (\bibinfo{year}{2014}{\natexlab{a}}).

\bibitem[{\citenamefont{Michieletto
  et~al.}(2014{\natexlab{b}})\citenamefont{Michieletto, Marenduzzo, Orlandini,
  Alexander, and Turner}}]{Turner2}
\bibinfo{author}{\bibfnamefont{D.}~\bibnamefont{Michieletto}},
  \bibinfo{author}{\bibfnamefont{D.}~\bibnamefont{Marenduzzo}},
  \bibinfo{author}{\bibfnamefont{E.}~\bibnamefont{Orlandini}},
  \bibinfo{author}{\bibfnamefont{G.~P.} \bibnamefont{Alexander}},
  \bibnamefont{and} \bibinfo{author}{\bibfnamefont{M.~S.}
  \bibnamefont{Turner}}, \bibinfo{journal}{Soft Matter}
  \textbf{\bibinfo{volume}{10}}, \bibinfo{pages}{5936}
  (\bibinfo{year}{2014}{\natexlab{b}}).

\bibitem[{\citenamefont{Michieletto and Turner}(2015)}]{Turner3}
\bibinfo{author}{\bibfnamefont{D.}~\bibnamefont{Michieletto}} \bibnamefont{and}
  \bibinfo{author}{\bibfnamefont{M.~S.} \bibnamefont{Turner}},
  \bibinfo{journal}{arXiv preprint arXiv:1510.05625}  (\bibinfo{year}{2015}).

\bibitem[{\citenamefont{Lee et~al.}(2015)\citenamefont{Lee, Kim, and
  Jung}}]{Threading4}
\bibinfo{author}{\bibfnamefont{E.}~\bibnamefont{Lee}},
  \bibinfo{author}{\bibfnamefont{S.}~\bibnamefont{Kim}}, \bibnamefont{and}
  \bibinfo{author}{\bibfnamefont{Y.}~\bibnamefont{Jung}},
  \bibinfo{journal}{Macromol. Rapid Commun.}  (\bibinfo{year}{2015}).

\bibitem[{\citenamefont{Everaers et~al.}(2004)\citenamefont{Everaers,
  Sukumaran, Grest, Svaneborg, Sivasubramanian, and Kremer}}]{Everaers}
\bibinfo{author}{\bibfnamefont{R.}~\bibnamefont{Everaers}},
  \bibinfo{author}{\bibfnamefont{S.~K.} \bibnamefont{Sukumaran}},
  \bibinfo{author}{\bibfnamefont{G.~S.} \bibnamefont{Grest}},
  \bibinfo{author}{\bibfnamefont{C.}~\bibnamefont{Svaneborg}},
  \bibinfo{author}{\bibfnamefont{A.}~\bibnamefont{Sivasubramanian}},
  \bibnamefont{and} \bibinfo{author}{\bibfnamefont{K.}~\bibnamefont{Kremer}},
  \bibinfo{journal}{Science} \textbf{\bibinfo{volume}{303}},
  \bibinfo{pages}{823} (\bibinfo{year}{2004}).

\bibitem[{\citenamefont{Kraft et~al.}(2013)\citenamefont{Kraft, Wittkowski, ten
  Hagen, Edmond, Pine, and L{\"o}wen}}]{ten_Hagen_Kraft_PRE_2013}
\bibinfo{author}{\bibfnamefont{D.~J.} \bibnamefont{Kraft}},
  \bibinfo{author}{\bibfnamefont{R.}~\bibnamefont{Wittkowski}},
  \bibinfo{author}{\bibfnamefont{B.}~\bibnamefont{ten Hagen}},
  \bibinfo{author}{\bibfnamefont{K.~V.} \bibnamefont{Edmond}},
  \bibinfo{author}{\bibfnamefont{D.~J.} \bibnamefont{Pine}}, \bibnamefont{and}
  \bibinfo{author}{\bibfnamefont{H.}~\bibnamefont{L{\"o}wen}},
  \bibinfo{journal}{Phys. Rev. E} \textbf{\bibinfo{volume}{88}},
  \bibinfo{pages}{050301} (\bibinfo{year}{2013}).

\bibitem[{\citenamefont{Gray and Gubbins}(1984)}]{GrayGubbins}
\bibinfo{author}{\bibfnamefont{C.}~\bibnamefont{Gray}} \bibnamefont{and}
  \bibinfo{author}{\bibfnamefont{K.}~\bibnamefont{Gubbins}},
  \emph{\bibinfo{title}{Theory of Molecular Fluids: I: Fundamentals}},
  International Series of Monographs on Chemistry (\bibinfo{publisher}{OUP
  Oxford}, \bibinfo{year}{1984}).

\bibitem[{\citenamefont{Onsager}(1949)}]{Onsager}
\bibinfo{author}{\bibfnamefont{L.}~\bibnamefont{Onsager}},
  \bibinfo{journal}{Ann. N.Y. Acad. Sci.} \textbf{\bibinfo{volume}{51}},
  \bibinfo{pages}{627} (\bibinfo{year}{1949}).

\bibitem[{\citenamefont{Lekkerkerker and Vroege}(2013)}]{Lekkerkerker}
\bibinfo{author}{\bibfnamefont{H.~N.~W.} \bibnamefont{Lekkerkerker}}
  \bibnamefont{and} \bibinfo{author}{\bibfnamefont{G.~J.}
  \bibnamefont{Vroege}}, \bibinfo{journal}{Phil. Trans. R. Soc. A}
  \textbf{\bibinfo{volume}{371}}, \bibinfo{pages}{20120263}
  (\bibinfo{year}{2013}).

\bibitem[{\citenamefont{Frenkel and Smit}(2001)}]{Book_Frenkel_Smit}
\bibinfo{author}{\bibfnamefont{D.}~\bibnamefont{Frenkel}} \bibnamefont{and}
  \bibinfo{author}{\bibfnamefont{B.}~\bibnamefont{Smit}},
  \emph{\bibinfo{title}{Understanding molecular simulation: from algorithms to
  applications}}, vol.~\bibinfo{volume}{1} (\bibinfo{publisher}{Academic
  press}, \bibinfo{year}{2001}).

\bibitem[{\citenamefont{Balberg}(1987)}]{balberg1987}
\bibinfo{author}{\bibfnamefont{I.}~\bibnamefont{Balberg}},
  \bibinfo{journal}{Philos. Mag. B} \textbf{\bibinfo{volume}{56}},
  \bibinfo{pages}{991} (\bibinfo{year}{1987}).

\bibitem[{\citenamefont{Cichocki and Hinsen}(1990)}]{Cichocki_Hinsen}
\bibinfo{author}{\bibfnamefont{B.}~\bibnamefont{Cichocki}} \bibnamefont{and}
  \bibinfo{author}{\bibfnamefont{K.}~\bibnamefont{Hinsen}},
  \bibinfo{journal}{Physica A} \textbf{\bibinfo{volume}{166}},
  \bibinfo{pages}{473} (\bibinfo{year}{1990}).

\bibitem[{\citenamefont{L{\"o}wen}(1994)}]{Loewen_PRE_1994}
\bibinfo{author}{\bibfnamefont{H.}~\bibnamefont{L{\"o}wen}},
  \bibinfo{journal}{Phys. Rev. E} \textbf{\bibinfo{volume}{50}},
  \bibinfo{pages}{1232} (\bibinfo{year}{1994}).

\bibitem[{\citenamefont{Scala et~al.}(2007)\citenamefont{Scala, Voigtmann, and
  De~Michele}}]{Sciortino}
\bibinfo{author}{\bibfnamefont{A.}~\bibnamefont{Scala}},
  \bibinfo{author}{\bibfnamefont{T.}~\bibnamefont{Voigtmann}},
  \bibnamefont{and}
  \bibinfo{author}{\bibfnamefont{C.}~\bibnamefont{De~Michele}},
  \bibinfo{journal}{J. Chem. Phys.} \textbf{\bibinfo{volume}{126}},
  \bibinfo{pages}{134109} (\bibinfo{year}{2007}).

\bibitem[{\citenamefont{Bra{\'n}ka and Heyes}(1998)}]{Heyes}
\bibinfo{author}{\bibfnamefont{A.}~\bibnamefont{Bra{\'n}ka}} \bibnamefont{and}
  \bibinfo{author}{\bibfnamefont{D.}~\bibnamefont{Heyes}},
  \bibinfo{journal}{Phys. Rev. E} \textbf{\bibinfo{volume}{58}},
  \bibinfo{pages}{2611} (\bibinfo{year}{1998}).

\bibitem[{\citenamefont{Strating}(1999)}]{Strating}
\bibinfo{author}{\bibfnamefont{P.}~\bibnamefont{Strating}},
  \bibinfo{journal}{Phys. Rev. E} \textbf{\bibinfo{volume}{59}},
  \bibinfo{pages}{2175} (\bibinfo{year}{1999}).

\bibitem[{\citenamefont{Kirchhoff et~al.}(1996)\citenamefont{Kirchhoff,
  L{\"o}wen, and Klein}}]{Kirchhoff_Klein_Loewen_1996}
\bibinfo{author}{\bibfnamefont{T.}~\bibnamefont{Kirchhoff}},
  \bibinfo{author}{\bibfnamefont{H.}~\bibnamefont{L{\"o}wen}},
  \bibnamefont{and} \bibinfo{author}{\bibfnamefont{R.}~\bibnamefont{Klein}},
  \bibinfo{journal}{Phys. Rev. E} \textbf{\bibinfo{volume}{53}},
  \bibinfo{pages}{5011} (\bibinfo{year}{1996}).

\bibitem[{\citenamefont{Wensink and L{\"o}wen}(2012)}]{Wensink_JPCM_2012}
\bibinfo{author}{\bibfnamefont{H.~H.} \bibnamefont{Wensink}} \bibnamefont{and}
  \bibinfo{author}{\bibfnamefont{H.}~\bibnamefont{L{\"o}wen}},
  \bibinfo{journal}{J. Phys. Condens. Matter} \textbf{\bibinfo{volume}{24}},
  \bibinfo{pages}{464130} (\bibinfo{year}{2012}).

\bibitem[{\citenamefont{Happel and Brenner}(1983)}]{happel_brenner}
\bibinfo{author}{\bibfnamefont{J.}~\bibnamefont{Happel}} \bibnamefont{and}
  \bibinfo{author}{\bibfnamefont{H.}~\bibnamefont{Brenner}},
  \emph{\bibinfo{title}{Low Reynolds number hydrodynamics: with special
  applications to particulate media}} (\bibinfo{publisher}{Springer Science \&
  Business Media}, \bibinfo{year}{1983}).

\bibitem[{\citenamefont{Fernandes and de~la Torre}(2002)}]{de_la_Torre}
\bibinfo{author}{\bibfnamefont{M.~X.} \bibnamefont{Fernandes}}
  \bibnamefont{and} \bibinfo{author}{\bibfnamefont{J.~G.} \bibnamefont{de~la
  Torre}}, \bibinfo{journal}{Biophys. J.} \textbf{\bibinfo{volume}{83}},
  \bibinfo{pages}{3039} (\bibinfo{year}{2002}).

\bibitem[{\citenamefont{Makino and Doi}(2004)}]{Doi}
\bibinfo{author}{\bibfnamefont{M.}~\bibnamefont{Makino}} \bibnamefont{and}
  \bibinfo{author}{\bibfnamefont{M.}~\bibnamefont{Doi}}, \bibinfo{journal}{J.
  Phys. Soc. Jpn.} \textbf{\bibinfo{volume}{73}}, \bibinfo{pages}{2739}
  (\bibinfo{year}{2004}).

\bibitem[{\citenamefont{Oberbeck}(1876)}]{oberbeck1876}
\bibinfo{author}{\bibfnamefont{A.}~\bibnamefont{Oberbeck}},
  \bibinfo{journal}{Journal f\"ur die reine und angewandte Mathematik}
  \textbf{\bibinfo{volume}{81}}, \bibinfo{pages}{62} (\bibinfo{year}{1876}).

\bibitem[{\citenamefont{Edwardes}(1893)}]{edwardes1893}
\bibinfo{author}{\bibfnamefont{D.}~\bibnamefont{Edwardes}},
  \bibinfo{journal}{The Quarterly Journal of Pure and Applied Mathematics}
  \textbf{\bibinfo{volume}{26}}, \bibinfo{pages}{260} (\bibinfo{year}{1893}).

\bibitem[{\citenamefont{Perrin}(1934)}]{perrin1934}
\bibinfo{author}{\bibfnamefont{F.}~\bibnamefont{Perrin}}, \bibinfo{journal}{J.
  Phys. Radium} \textbf{\bibinfo{volume}{5}}, \bibinfo{pages}{497}
  (\bibinfo{year}{1934}).

\bibitem[{\citenamefont{Passow et~al.}(2015)\citenamefont{Passow, ten Hagen,
  L{\"o}wen, and Wagner}}]{CWagner}
\bibinfo{author}{\bibfnamefont{C.}~\bibnamefont{Passow}},
  \bibinfo{author}{\bibfnamefont{B.}~\bibnamefont{ten Hagen}},
  \bibinfo{author}{\bibfnamefont{H.}~\bibnamefont{L{\"o}wen}},
  \bibnamefont{and} \bibinfo{author}{\bibfnamefont{J.}~\bibnamefont{Wagner}},
  \bibinfo{journal}{J. Chem. Phys.} \textbf{\bibinfo{volume}{143}},
  \bibinfo{pages}{044903} (\bibinfo{year}{2015}).

\bibitem[{\citenamefont{Ermak}(1975)}]{Ermak_1975_JCP}
\bibinfo{author}{\bibfnamefont{D.~L.} \bibnamefont{Ermak}},
  \bibinfo{journal}{J. Chem. Phys.} \textbf{\bibinfo{volume}{62}},
  \bibinfo{pages}{4189} (\bibinfo{year}{1975}).

\bibitem[{\citenamefont{Beard and Schlick}(2003)}]{biasfreerotation}
\bibinfo{author}{\bibfnamefont{D.~A.} \bibnamefont{Beard}} \bibnamefont{and}
  \bibinfo{author}{\bibfnamefont{T.}~\bibnamefont{Schlick}},
  \bibinfo{journal}{Biophys. J.} \textbf{\bibinfo{volume}{85}},
  \bibinfo{pages}{2973} (\bibinfo{year}{2003}).

\bibitem[{\citenamefont{Lorenz and Ziff}(1998)}]{lorenz1998}
\bibinfo{author}{\bibfnamefont{C.~D.} \bibnamefont{Lorenz}} \bibnamefont{and}
  \bibinfo{author}{\bibfnamefont{R.~M.} \bibnamefont{Ziff}},
  \bibinfo{journal}{Phys. Rev. E} \textbf{\bibinfo{volume}{57}},
  \bibinfo{pages}{230} (\bibinfo{year}{1998}).

\bibitem[{\citenamefont{Newman}(2003)}]{newman_networks}
\bibinfo{author}{\bibfnamefont{M.~E.~J.} \bibnamefont{Newman}},
  \bibinfo{journal}{SIAM Rev.} \textbf{\bibinfo{volume}{45}},
  \bibinfo{pages}{167} (\bibinfo{year}{2003}).

\bibitem[{\citenamefont{Albert and Barab{\'a}si}(2002)}]{albert_networks}
\bibinfo{author}{\bibfnamefont{R.}~\bibnamefont{Albert}} \bibnamefont{and}
  \bibinfo{author}{\bibfnamefont{A.-L.} \bibnamefont{Barab{\'a}si}},
  \bibinfo{journal}{Rev. Mod. Phys.} \textbf{\bibinfo{volume}{74}},
  \bibinfo{pages}{47} (\bibinfo{year}{2002}).

\bibitem[{\citenamefont{L\"owen et~al.}(1993)\citenamefont{L\"owen, Palberg,
  and Simon}}]{lps1993}
\bibinfo{author}{\bibfnamefont{H.}~\bibnamefont{L\"owen}},
  \bibinfo{author}{\bibfnamefont{T.}~\bibnamefont{Palberg}}, \bibnamefont{and}
  \bibinfo{author}{\bibfnamefont{R.}~\bibnamefont{Simon}},
  \bibinfo{journal}{Phys. Rev. Lett.} \textbf{\bibinfo{volume}{70}},
  \bibinfo{pages}{1557} (\bibinfo{year}{1993}).

\bibitem[{\citenamefont{L\"owen and Szamel}(1993)}]{Szamel}
\bibinfo{author}{\bibfnamefont{H.}~\bibnamefont{L\"owen}} \bibnamefont{and}
  \bibinfo{author}{\bibfnamefont{G.}~\bibnamefont{Szamel}},
  \bibinfo{journal}{J. Phys. Condens. Matter} \textbf{\bibinfo{volume}{5}},
  \bibinfo{pages}{2295} (\bibinfo{year}{1993}).

\bibitem[{\citenamefont{L{\"o}wen}(1996)}]{2dLinde}
\bibinfo{author}{\bibfnamefont{H.}~\bibnamefont{L{\"o}wen}},
  \bibinfo{journal}{Phys. Rev. E} \textbf{\bibinfo{volume}{53}},
  \bibinfo{pages}{R29} (\bibinfo{year}{1996}).

\bibitem[{\citenamefont{Dunleavy et~al.}(2012)\citenamefont{Dunleavy, Wiesner,
  and Royall}}]{Wiesner}
\bibinfo{author}{\bibfnamefont{A.~J.} \bibnamefont{Dunleavy}},
  \bibinfo{author}{\bibfnamefont{K.}~\bibnamefont{Wiesner}}, \bibnamefont{and}
  \bibinfo{author}{\bibfnamefont{C.~P.} \bibnamefont{Royall}},
  \bibinfo{journal}{Phys. Rev. E} \textbf{\bibinfo{volume}{86}},
  \bibinfo{pages}{041505} (\bibinfo{year}{2012}).

\bibitem[{\citenamefont{Williams et~al.}(2015)\citenamefont{Williams,
  O{\u{g}}uz, Bartlett, L{\"o}wen, and Royall}}]{Williams}
\bibinfo{author}{\bibfnamefont{I.}~\bibnamefont{Williams}},
  \bibinfo{author}{\bibfnamefont{E.~C.} \bibnamefont{O{\u{g}}uz}},
  \bibinfo{author}{\bibfnamefont{P.}~\bibnamefont{Bartlett}},
  \bibinfo{author}{\bibfnamefont{H.}~\bibnamefont{L{\"o}wen}},
  \bibnamefont{and} \bibinfo{author}{\bibfnamefont{C.~P.}
  \bibnamefont{Royall}}, \bibinfo{journal}{J. Chem. Phys.}
  \textbf{\bibinfo{volume}{142}}, \bibinfo{pages}{024505}
  (\bibinfo{year}{2015}).

\bibitem[{\citenamefont{Angelani et~al.}(1998)\citenamefont{Angelani, Parisi,
  Ruocco, and Viliani}}]{Angelani}
\bibinfo{author}{\bibfnamefont{L.}~\bibnamefont{Angelani}},
  \bibinfo{author}{\bibfnamefont{G.}~\bibnamefont{Parisi}},
  \bibinfo{author}{\bibfnamefont{G.}~\bibnamefont{Ruocco}}, \bibnamefont{and}
  \bibinfo{author}{\bibfnamefont{G.}~\bibnamefont{Viliani}},
  \bibinfo{journal}{Phys. Rev. Lett.} \textbf{\bibinfo{volume}{81}},
  \bibinfo{pages}{4648} (\bibinfo{year}{1998}).

\bibitem[{\citenamefont{Rabani et~al.}(1999)\citenamefont{Rabani, Gezelter, and
  Berne}}]{Rabani}
\bibinfo{author}{\bibfnamefont{E.}~\bibnamefont{Rabani}},
  \bibinfo{author}{\bibfnamefont{J.~D.} \bibnamefont{Gezelter}},
  \bibnamefont{and} \bibinfo{author}{\bibfnamefont{B.~J.} \bibnamefont{Berne}},
  \bibinfo{journal}{Phys. Rev. Lett.} \textbf{\bibinfo{volume}{82}},
  \bibinfo{pages}{3649} (\bibinfo{year}{1999}).

\bibitem[{\citenamefont{G{\"o}tze}(2008)}]{Goetze_book}
\bibinfo{author}{\bibfnamefont{W.}~\bibnamefont{G{\"o}tze}},
  \emph{\bibinfo{title}{Complex Dynamics of Glass-Forming Liquids: A
  Mode-Coupling Theory}}, vol. \bibinfo{volume}{143}
  (\bibinfo{publisher}{Oxford University Press}, \bibinfo{year}{2008}).

\bibitem[{\citenamefont{Puertas and Voigtmann}(2014)}]{puertas_microrheology}
\bibinfo{author}{\bibfnamefont{A.~M.} \bibnamefont{Puertas}} \bibnamefont{and}
  \bibinfo{author}{\bibfnamefont{T.}~\bibnamefont{Voigtmann}},
  \bibinfo{journal}{J. Phys. Condens. Matter} \textbf{\bibinfo{volume}{26}},
  \bibinfo{pages}{243101} (\bibinfo{year}{2014}).

\bibitem[{\citenamefont{Carnevale et~al.}(1990)\citenamefont{Carnevale, Pomeau,
  and Young}}]{Carnevale}
\bibinfo{author}{\bibfnamefont{G.~F.} \bibnamefont{Carnevale}},
  \bibinfo{author}{\bibfnamefont{Y.}~\bibnamefont{Pomeau}}, \bibnamefont{and}
  \bibinfo{author}{\bibfnamefont{W.~R.} \bibnamefont{Young}},
  \bibinfo{journal}{Phys. Rev. Lett.} \textbf{\bibinfo{volume}{64}},
  \bibinfo{pages}{2913} (\bibinfo{year}{1990}).

\bibitem[{\citenamefont{Kolb}(1984)}]{Kolb}
\bibinfo{author}{\bibfnamefont{M.}~\bibnamefont{Kolb}}, \bibinfo{journal}{Phys.
  Rev. Lett.} \textbf{\bibinfo{volume}{53}}, \bibinfo{pages}{1653}
  (\bibinfo{year}{1984}).

\bibitem[{\citenamefont{Trizac and Hansen}(1996)}]{Hansen}
\bibinfo{author}{\bibfnamefont{E.}~\bibnamefont{Trizac}} \bibnamefont{and}
  \bibinfo{author}{\bibfnamefont{J.-P.} \bibnamefont{Hansen}},
  \bibinfo{journal}{J. Stat. Phys.} \textbf{\bibinfo{volume}{82}},
  \bibinfo{pages}{1345} (\bibinfo{year}{1996}).

\bibitem[{\citenamefont{Wensink and L{\"o}wen}(2006)}]{Wensink}
\bibinfo{author}{\bibfnamefont{H.~H.} \bibnamefont{Wensink}} \bibnamefont{and}
  \bibinfo{author}{\bibfnamefont{H.}~\bibnamefont{L{\"o}wen}},
  \bibinfo{journal}{Phys. Rev. Lett.} \textbf{\bibinfo{volume}{97}},
  \bibinfo{pages}{038303} (\bibinfo{year}{2006}).

\bibitem[{\citenamefont{Cremer and L{\"o}wen}(2014)}]{Cremer}
\bibinfo{author}{\bibfnamefont{P.}~\bibnamefont{Cremer}} \bibnamefont{and}
  \bibinfo{author}{\bibfnamefont{H.}~\bibnamefont{L{\"o}wen}},
  \bibinfo{journal}{Phys. Rev. E} \textbf{\bibinfo{volume}{89}},
  \bibinfo{pages}{022307} (\bibinfo{year}{2014}).

\bibitem[{\citenamefont{Harth et~al.}(2013)\citenamefont{Harth, Kornek,
  Trittel, Strachauer, H{\"o}me, Will, and Stannarius}}]{Stannarius}
\bibinfo{author}{\bibfnamefont{K.}~\bibnamefont{Harth}},
  \bibinfo{author}{\bibfnamefont{U.}~\bibnamefont{Kornek}},
  \bibinfo{author}{\bibfnamefont{T.}~\bibnamefont{Trittel}},
  \bibinfo{author}{\bibfnamefont{U.}~\bibnamefont{Strachauer}},
  \bibinfo{author}{\bibfnamefont{S.}~\bibnamefont{H{\"o}me}},
  \bibinfo{author}{\bibfnamefont{K.}~\bibnamefont{Will}}, \bibnamefont{and}
  \bibinfo{author}{\bibfnamefont{R.}~\bibnamefont{Stannarius}},
  \bibinfo{journal}{Phys. Rev. Lett.} \textbf{\bibinfo{volume}{110}},
  \bibinfo{pages}{144102} (\bibinfo{year}{2013}).

\bibitem[{\citenamefont{K{\"u}mmel et~al.}(2013)\citenamefont{K{\"u}mmel, ten
  Hagen, Wittkowski, Buttinoni, Eichhorn, Volpe, L{\"o}wen, and
  Bechinger}}]{Bechinger1}
\bibinfo{author}{\bibfnamefont{F.}~\bibnamefont{K{\"u}mmel}},
  \bibinfo{author}{\bibfnamefont{B.}~\bibnamefont{ten Hagen}},
  \bibinfo{author}{\bibfnamefont{R.}~\bibnamefont{Wittkowski}},
  \bibinfo{author}{\bibfnamefont{I.}~\bibnamefont{Buttinoni}},
  \bibinfo{author}{\bibfnamefont{R.}~\bibnamefont{Eichhorn}},
  \bibinfo{author}{\bibfnamefont{G.}~\bibnamefont{Volpe}},
  \bibinfo{author}{\bibfnamefont{H.}~\bibnamefont{L{\"o}wen}},
  \bibnamefont{and}
  \bibinfo{author}{\bibfnamefont{C.}~\bibnamefont{Bechinger}},
  \bibinfo{journal}{Phys. Rev. Lett.} \textbf{\bibinfo{volume}{110}},
  \bibinfo{pages}{198302} (\bibinfo{year}{2013}).

\bibitem[{\citenamefont{ten Hagen et~al.}(2014)\citenamefont{ten Hagen,
  K{\"u}mmel, Wittkowski, Takagi, L{\"o}wen, and Bechinger}}]{Bechinger2}
\bibinfo{author}{\bibfnamefont{B.}~\bibnamefont{ten Hagen}},
  \bibinfo{author}{\bibfnamefont{F.}~\bibnamefont{K{\"u}mmel}},
  \bibinfo{author}{\bibfnamefont{R.}~\bibnamefont{Wittkowski}},
  \bibinfo{author}{\bibfnamefont{D.}~\bibnamefont{Takagi}},
  \bibinfo{author}{\bibfnamefont{H.}~\bibnamefont{L{\"o}wen}},
  \bibnamefont{and}
  \bibinfo{author}{\bibfnamefont{C.}~\bibnamefont{Bechinger}},
  \bibinfo{journal}{Nat. Commun.} \textbf{\bibinfo{volume}{5}}
  (\bibinfo{year}{2014}).

\bibitem[{\citenamefont{Iwaki et~al.}(2015)\citenamefont{Iwaki, Ishido, Hirano,
  Lazutin, Vasilevskaya, Kenmotsu, and Yoshikawa}}]{Yoshikawa}
\bibinfo{author}{\bibfnamefont{T.}~\bibnamefont{Iwaki}},
  \bibinfo{author}{\bibfnamefont{T.}~\bibnamefont{Ishido}},
  \bibinfo{author}{\bibfnamefont{K.}~\bibnamefont{Hirano}},
  \bibinfo{author}{\bibfnamefont{A.~A.} \bibnamefont{Lazutin}},
  \bibinfo{author}{\bibfnamefont{V.~V.} \bibnamefont{Vasilevskaya}},
  \bibinfo{author}{\bibfnamefont{T.}~\bibnamefont{Kenmotsu}}, \bibnamefont{and}
  \bibinfo{author}{\bibfnamefont{K.}~\bibnamefont{Yoshikawa}},
  \bibinfo{journal}{J. Chem. Phys.} \textbf{\bibinfo{volume}{142}},
  \bibinfo{pages}{145101} (\bibinfo{year}{2015}).

\end{thebibliography}

\end{document}